\documentclass[prc,twocolumn,nofootinbib,superscriptaddress,showpacs]{revtex4}
\usepackage{graphicx,amsmath,amssymb,bm,multirow}
\usepackage{color}
\usepackage{amscd}

\newcommand{\be}[1]{\begin{equation}\label{#1}}
\newcommand{\ee}{\end{equation}}
\newcommand{\vlowk}{V_{{\rm low}\,k}}

\newcommand{\fmicube}{\, \text{fm}^{-3}}
\newcommand{\mev}{\, \text{MeV}}

\begin{document}

\title{Superfluid properties of the inner crust of neutron stars II. Wigner-Seitz cells at finite temperature}

\author{Alessandro Pastore}
\email{pastore@inpl.in2p3.fr}
\affiliation{Universit\'e de Lyon, F-69003 Lyon, France; Universit\'e Lyon 1, 
43 Bd. du 11 Novembre 1918, F-69622 Villeurbanne
cedex, France CNRS-IN2P3, UMR 5822, Institut de Physique Nucl\'{e}aire de Lyon}

\date{\today}

\begin{abstract}
We investigate the superfluid properties of the inner crust of neutron stars at finite temperature for different pairing functionals.
We generalize the formalism adopted in  article \cite{Pastore11} to include the effect of the temperature to calculate the specific heat of each given Wigner-Seitz cell.
The calculations are done for two  pairing forces, Gogny D1 and $V_{\text{low-k}}$, with finite range and a density dependent contact interaction. We compare in such a way the effect of the pairing strength and of the range on the thermal properties of the inner crust.
\end{abstract}

\pacs{+++}

\maketitle

\section{\label{sec:intro}Introduction}

The study of  thermal properties of the surface of neutron stars represents a possible way to provide strong constraints on its inner structure.
In particular some authors~\cite{Lattimer} have argued that the occurrence of rapid cooling processes in the interior of the neutron star and the observation of the variations of the surface temperature can give a possible constraint on the equation of state (EoS) of the dense matter.
It follows that it is important to obtain a precise model describing the cooling mechanism of the star. 

It has been shown~\cite{Petick,Petick2} that the cooling time of the neutron star is mainly determined by the thermal response of the inner crust.
This region of the star is formed by nuclear clusters surrounded by a sea of unbound neutrons and ultra-relativistic electrons (see ref. \cite{Chamel_2008} and references therein).
Thus the importance of performing accurate calculations  of its specific heat, $C_{V}$.
Within the literature, it exists already  several calculations for the inner crust and for its thermal properties.  
In a recent article, Baldo \emph{et al.}~\cite{Baldo_2006} has shown that pairing correlations play an important role for this region of the neutron star. Including such correlations the sequence of Wigner-Seitz (WS) cells is remarkably modified compared to the original one proposed by Negele and Vautherin~\cite{Negele_1973}.
Several authors have  done calculations in this system using non-relativistic functionals and with zero-range pairing interactions~\cite{Grill_2011,Sandulescu_2004,Chamel10,Chamel09,Montani04} or finite range pairing interactions~\cite{Pizzochero_2002,Barranco_1998}, but without performing a systematic comparison among the two.

For this reason, we decided to continue the investigation we started in our previous article~\cite{Pastore11} (hereafter called article I) on the properties of different pairing functionals at finite temperature.
In particular we decided to study three different pairing interactions: two  with finite range ($i.e.$ $V_{\text{low-k}}$ and Gogny D1), but with different  strengths and one zero-range  fitted to reproduce the properties of the finite range one in case of infinite  systems.
This  allow us to better comprehend the role of the range of the interaction and of the strength in our results.
\begin{table}
\begin{center}
\begin{tabular}{ccccccc}
\hline
\hline
Zone & Z & N& $R_{WS}$ [fm] & $\rho_n ^{b}$ &$k_{F}^{n}$ [fm$^{-1}$] &$\bar{k}_{F}^{n}$ [fm$^{-1}$] \\
\hline
11 & 40 & 140 & 53.6 & $7.93\times 10^{-5}$ & 0.13&0.18\\
10 & 40 & 160 & 49.2 & $1.38\times10^{-4}$ & 0.16 &0.21\\
 9 &  40 & 210 & 46.4 & $2.78\times 10^{-4}$ & 0.20 & 0.25\\
 8 & 40 & 280 & 44.4 & $5.02\times 10^{-4}$ & 0.24  &0.28\\
 7 &  40 & 460 & 42.2 & $1.15\times10^{-3}$ & 0.32 & 0.35\\
 6 &  50 & 900 & 39.3 & $2.99\times 10^{-3}$ & 0.44 & 0.47\\
 5 &  50 & 1050 & 35.7 & $4.75\times10^{-3}$ &0.52 & 0.55\\
 4 &  50 & 1300 & 33.0 & $7.54\times 10^{-3}$  &0.61&0.63 \\
 3 &  50 & 1750 & 27.6 & $1.77\times10^{-2}$  &0.81& 0.85\\
 2 &  40 & 1460 & 19.6 & $4.23\times 10^{-2}$  &1.08&1.14 \\
 1 &  32 & 950 & 14.4 & - & -&1.37 \\
\hline
\end{tabular}
\caption{The WS cells representing different density regions of the inner crust and used in the present study. 
In the different columns we have: the particle numbers Z, N, the WS cell radii
$R_{WS}$, the background density $\rho_n^{b}$ (obtained  averaging the  neutron gas density far away from the cluster) and its  Fermi momentum $k_{F}^{n}$. In the last column we also give  $\bar{k}_{F}^{n}$, which  is the Fermi momentum calculated using   the average neutron density of the WS cell.}
\label{tabWS}
\end{center}
\end{table}

As done in article I, we adopt the Wigner-Seitz  approximation to study the different regions of the inner crust. For the validity of this approximation we address to the article of Chamel \emph{et al.}~\cite{Chamel07}.
Within each WS cell we perform fully self-consistent Hartree-Fock-Bogoliubov (HFB) calculations, in this way we can obtain the entropy and the specific heat of each cell as a function of the temperature for both protons and neutrons.

The article is organized as follow in Sec.\ref{Sect:calc_details}, we present the formalism of our calculations in both Wigner-Seitz cells and pure neutron matter (PNM), in Sec.\ref{Sect:results} we discuss the results we obtained concerning the densities, pairing gaps and specific heats and finally in Sec. \ref{Sect:conclusions}, we present our conclusions.

\section{Calculation details} \label{Sect:calc_details}

\subsection{Inner crust of neutron stars}\label{Subsect:innercrust}

\begin{figure}
\begin{center}
\includegraphics[clip=,angle=-90,width=0.48\textwidth]{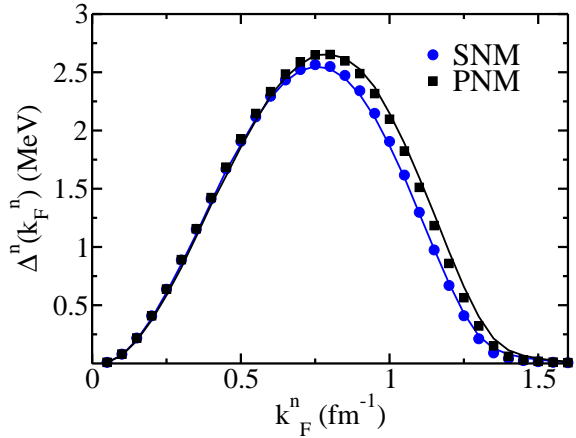}
\end{center}
\caption{
(Colors online)We compare the value of the neutron pairing gap at the Fermi energy in symmetric and pure neutron matter calculated using the $V_{\text{low-k}}$ (squares and circles respectively) and the DDDI interaction defined in Eq.\ref{pairing_int_contact} (solid lines). See text for details. }
\label{FitDDDI}
\end{figure}

We assume that the cluster structure of the inner crust is the one given by Negele and Vautherin \cite{Negele_1973}. We summarize the results in Tab.\ref{tabWS}.
The cells are assumed to be spherical on the average and non-interacting among each others.

The HFB equations are extended to include the effects of finite temperature and they take the form~\cite{Goodman86}

\begin{eqnarray}\label{HFBeq}
  \sum_{n'}(\bar{h}_{n'nlj}^q- \mu_{F,q})U^{i,q}_{n'lj}+\sum_{n'}\bar{\Delta}_{nn'lj}^qV^{i,q}_{n'lj}=E^{q}_{ilj}U^{i,q}_{nlj}, & & 
     \nonumber \\
  \sum_{n'}\bar{\Delta}^q_{nn'lj}U^{i,q}_{n'lj} -\sum_{n'}(\bar{h}^q_{n'nlj}- \mu_{F,q})V^{i,q}_{n'lj}  =E^{q}_{ilj}V^{i,q}_{nlj}. & & 
      \nonumber \\
\end{eqnarray}
Where  $q$ stands for neutrons (n) and protons (p); with $\mu_{F,q}$ we indicate the chemical potential (also called Fermi energy)\cite{Ring_1980}. The latter plays the role of a Lagrange multiplier to constraint on average the number of particles for each species separately. 
We used the standard notation $nlj$ for the spherical single-particle
states with radial quantum number $n$, orbital angular momentum $l$ and
total angular momentum $j$. 
$U^{i,q}_{nlj}$ and $V^{i,q}_{nlj}$ are the Bogoliubov amplitudes for the $i$-th quasiparticle
of energy $E^{q}_{ilj}$.
 $\bar{h}_{n'nlj}^q$ is the thermal averaged single particle hamiltonian calculated using a Skyrme functional.
 In this case, as illustrated by Bonche \emph{et al.}~\cite{Bonche84,Bonche85}, the structure of $\bar{h}_{n'nlj}^q$ is formally the same at zero temperature, what changes is the structure of local densities that now include an additional Fermi-Dirac distribution. The explicit expression of such quantities can be found in ref.~\cite{Sandulescu_2004}.
 
 The HFB equations are solved in a spherical box using the Dirichlet-Neumann mixed boundary conditions   as done in the original Negele and Vautherin's article~\cite{Negele_1973}. As indicated in article I we have two choise for the boundary conditions : (i) even-parity wave functions vanish at $R=R_{WS}$; (ii) the first derivative of odd-parity wave functions vanishes at $R=R_{WS}$. We call them Boundary Conditions Even (BCE), while Boundary Condition Odd (BCO) when we do the opposite.
In the rest of the article we will always use the BCE condition, unless when explicitly indicated the BCO ones. 
For the $ph$ channel we adopt the Skyrme functional SLy4~\cite{Chabanat97,Chabanat98}. The dependence of the results on the choice of the Skyrme functional has been already discussed in article I.
For the $pp$ channel we consider three different two-body pairing interactions: (i) a density-dependent 
contact interaction (DDDI)~\cite{Bertsch_1991}; (ii) the finite-range Gogny D1 interaction~\cite{Tian_2009a}; (iii) low-momentum
realistic interactions ($\vlowk$)~\cite{Duguet_2004,Lesisnki_2008,Lesinski_2009,Duguet_2009}.
More details about the finite range pairing interactions have been already given in article I.

According to our previous results, we found that the simple DDDI, adopted in article I, is not adequate to treat on equal footing both neutron and proton superfluidity.
Following Margueron \emph{et al.}~\cite{Margueron07,Margueron08}, we introduce a more complicated form for the contact interaction  

\begin{widetext}
\begin{eqnarray}\label{pairing_int_contact}
\qquad \quad v^{q}(\mathbf{r}_{1},\mathbf{r}_{2})=V_{0}\left[ 1- f^{q}_{1}\eta_{1} \left( \frac{\rho_0\left(
\frac{\mathbf{r}_{1}+\mathbf{r}_{2}}{2}\right)}{\rho_{sat}}\right)^{\alpha_{1}}-f^{q}_{2} \eta_{2} \left( \frac{\rho_0\left(
\frac{\mathbf{r}_{1}+\mathbf{r}_{2}}{2}\right)}{\rho_{sat}}\right)^{\alpha_{2}}\right]
\delta(\mathbf{r}_{1}-\mathbf{r}_{2}), & & \nonumber \\
& &
\end{eqnarray} 
\end{widetext}

\noindent where $\rho_{0}(r)$ is the isoscalar density and $\rho_{sat}=0.16 \fmicube$  is the saturation density .
The parameters $f_{1}^{q},f_{2}^{q}$ are defined as

\begin{eqnarray}
\begin{matrix}
f_{1}^{n}=\frac{\rho_{n}-\rho_{p}}{\rho_{n}+\rho_{p}}, & f_{2}^{n}=1-f^{n}_{1},\\
f_{1}^{p}=\frac{\rho_{p}-\rho_{n}}{\rho_{n}+\rho_{p}}, & f_{2}^{p}=1-f^{p}_{1}.
\end{matrix}
\end{eqnarray}

To avoid the utraviolet divergency we introduce a sharp cut-off of $60\mev$ on the quasiparticle energy~\cite{Bulgac}.
The strength of the interaction, $V_{0}=-458.0$ MeV fm$^3$, is related to the cut off via the standard relation~\cite{Garrido}

\begin{equation}
V_{0}=-\frac{2\pi^{2}\hbar^{2}m^{-1}}{\sqrt{\frac{2m}{\hbar^{2}E_{cut}}}-\frac{\pi}{2a_{nn}}},
\end{equation}

\noindent where $a_{nn}=-18.5$ fm is the neutron-neutron scattering length. The parameters $\eta_{1}=0.71,\eta_{2}=1.03$, $\alpha_{1}=0.51,\alpha_{2}=0.51$ are fixed to reproduce the results obtained with $V_{\text{low-k}}$ pairing interaction in both Symmetric Nuclear Matter (SNM) and Pure Neutron Matter  using the SLy4 mean field.
In Fig.\ref{FitDDDI} we show the results of our fitting procedure. We notice that  with our choice of parameters $\eta_{i},\alpha_{i}$ the DDDI gives a satisfactory description of the pairing gaps 
  of the $V_{\text{low-k}}$ interaction at the Fermi energy.
  It is important to recall that the DDDI interaction can reproduce the matrix elements of the $V_{\text{low-k}}$ interaction only at the Fermi energy, we refer to the discussion in article~\cite{Duguet_2004} for more details.

As done in article I, we drop the Coulomb term in the proton-proton pairing channel, this leads to an overestimate of about 10\% of the proton pairing gaps.

To have a better comparison with the existing literature, in this work we include also the proton-electron interaction \cite{Bonche81} that reads

\begin{equation}
V^{pe}(r)=\frac{Ze^{2}}{2R_{WS}}\left[ \left(\frac{r}{R_{WS}}\right)^{2}-3\right].
\end{equation}

\noindent This equation is valid under the assumption  of uniform electron distribution.
\noindent When presenting the results for the HFB  pairing gaps, we
show the Lowest-quasiparticle-energy Canonical State (LCS) pairing gaps~\cite{Lesisnki_2008,Duguet_2009}.
It is defined as the diagonal pairing matrix element corresponding to the canonical single-particle state, whose quasi-particle energy 

\begin{equation}\label{gapLCS}
E^{q}_{nlj}=\sqrt{\left( \varepsilon^{q}_{nlj}-\varepsilon_{F,q} \right)^{2}+\left(\Delta_{nlj}^{q}\right)^{2}}
\end{equation}

\noindent is the lowest. Here $\varepsilon^{q}_{nlj}$ stands for the diagonal matrix element of the single-particle field $h^{q}$ in canonical basis and $\Delta_{nlj}^{q}$ the corresponding diagonal pairing-field matrix element.

\subsection{Pure neutron matter}\label{pureN}
\begin{figure*}
\begin{center}
\includegraphics[clip=,angle=-90,width=0.45\textwidth]{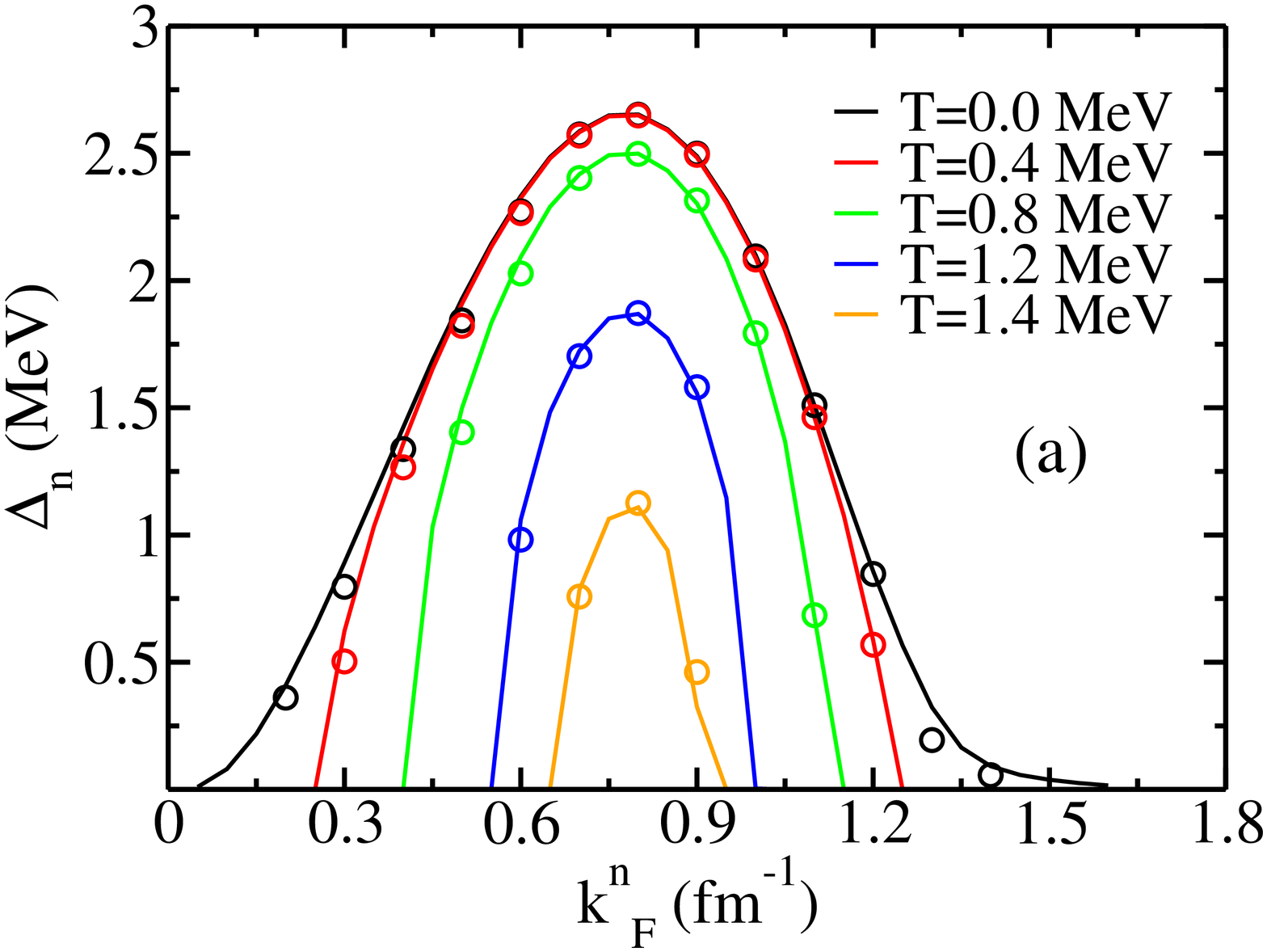}
\includegraphics[clip=,angle=-90,width=0.45\textwidth]{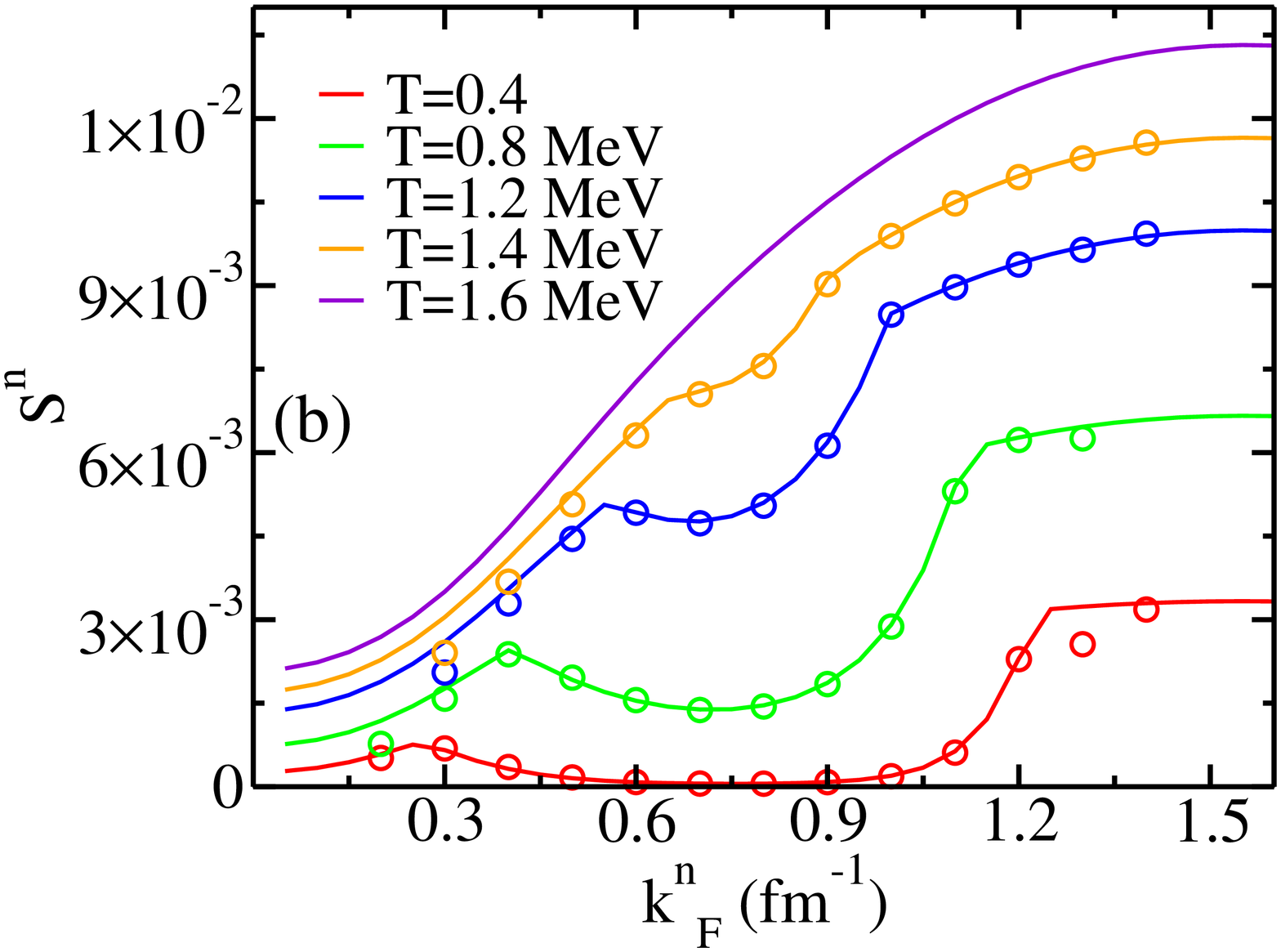}
\end{center}
\caption{
(Colors online) On the left, we show the neutron pairing gap Eq.\ref{gapeqINM} in PNM for the $V_{low-k}$ pairing interaction at different temperatures.  On the right we show the density of entropy of the system as a function of the neutron Fermi momentum, $k_{F}^{n}$.
In both cases the solid lines correspond to the solutions of Eq.\ref{gapeqINM}, while the dots are the solution of Eq.\ref{HFBeq}. See text for details.}
\label{gapPNMT}
\end{figure*}

\begin{figure}
\begin{center}
\includegraphics[clip=,angle=-90,width=0.45\textwidth]{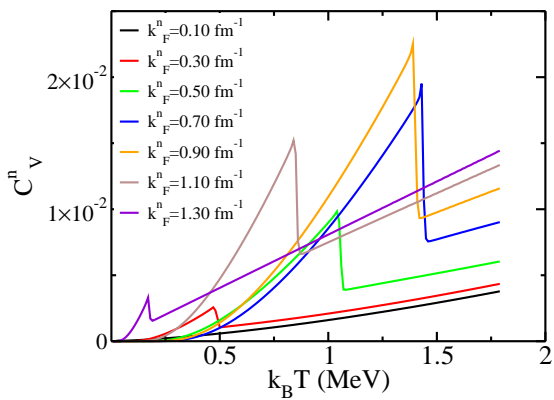}
\end{center}
\caption{
(Colors online) Specific heat $C_{V}^{n}$ of neutrons, Eq.\ref{spec_heatPNM}, calculated in PNM performing the derivative of the entropy calculated using Eq.~\ref{EqentropyPNM}. See text for details.}
\label{CVPNM}
\end{figure}

Before studying the effect of the presence of bound nucleons in the center of the WS cell,
we analyze the superfluid properties of the  pure neutron matter at finite temperature.
For a given neutron density $\rho_n$ in PNM, the gap and number equations for a given temperature $T$,
have to be solved simultaneously \cite{Ring89}
\begin{equation}\label{gapeqINM}
\Delta_n(k)=-\frac{1}{2}\int \frac{d^3k'}{(2\pi)^3}v(k-k')\frac{\Delta_n(k')}{E_n(k')}\tanh\left( \frac{E_n(k')}{2k_{B}T}\right)
\end{equation}
\begin{equation}\label{numbeqINM}
\rho_n=\frac{1}{2\pi^2}\int dk \; k^2
                \left[
                 1-\frac{\varepsilon_n(k)-\mu_n}{E_n(k)}\tanh\left( \frac{E_n(k')}{2k_{B}T}\right)
                \right] .
\end{equation}
$\mu_n$ is the neutron chemical potential, $k_{B}$ is the Boltzman constant and $E_n(k)=\sqrt{(\varepsilon_n(k)-\mu_n)^2+\Delta_n(k)^2}$ 
is the quasiparticle energy, while the single-particle energy $\varepsilon_n(k)$ is given by
the sum of the kinetic energy and the Hartree-Fock potential $\bar{U}^n_{HF}$
\begin{equation}\label{spINM}
  \varepsilon_n(k)=\frac{\hbar^2k^2}{2m_n^*} + \bar{U}^n_{HF}(k) .
\end{equation}
Where $m_n^*$ is the neutron effective mass.
The temperature dependence is introduced through the term $\tanh\left( \frac{E_n(k')}{2k_{B}T}\right)$.
The number equation (cf. Eq.~(\ref{numbeqINM})) provides the relation between the density 
and the chemical potential $\mu_n$. In the limit of weak coupling, where $\Delta_n << \varepsilon_{F,n}$,
the chemical potential can be approximated by the Fermi energy $\varepsilon_{F,n}=\frac{\hbar^2 \left(k_{F}^{n}\right)^{2}}{2m_n^*}$,
with $k_{F}^{n}=(3\pi^2\rho_n)^{1/3}$. 
The entropy of the infinite system for unit volume is~\cite{Grill_2011}

\begin{equation}\label{EqentropyPNM}
S^{n}=g \int \frac{d^{3}k}{(2\pi)^{3}} [f^{n}(k)\ln f^{n}(k)+(1-f^{n}(k))\ln (1-f^{n}(k))]
\end{equation}

\noindent where $f^{n}(k)=\left( 1+\exp \frac{E_{n}(k)}{k_{B}T}\right)^{-1}$ is the Fermi distribution, and $g=2$ is the degeneracy of the system.

\noindent To test the accuracy of our code we compare in Fig.\ref{gapPNMT} (a)  the solution of the infinite system using Eq.\ref{gapeqINM} and using the HFB equations defined in Eq.\ref{HFBeq}, for a system without protons.
The HFB equations, Eq.\ref{HFBeq},  are solved in a box of 35 fm with the BCE conditions.
The test has been done adopting the $V_{\text{low-k}}$ pairing interaction together with the Skyrme functional SLy4, that determines the neutron effective mass, $m_{n}^{*}$.
We observe that the two methods are numerically equivalent.
In  Fig.\ref{gapPNMT} (b) we show the entropy of the systems as a function of the neutron Fermi momentum, $k_{F}^{n}$, calculated with the two methods.
We notice  again that the two techniques used to describe the PNM system  are numerically equivalent. This remains true also for the others pairing interactions.

It is possible now to define the specific heat per unit of volume as

\begin{equation}\label{spec_heatPNM}
C_{V}^{n}=T\frac{d S^{n}}{dT}.
\end{equation}

\noindent The result is shown in  Fig.\ref{CVPNM}  for different values of the Fermi momentum as a function of the temperature. The specific heat increases as a function of the temperature, up to a given critical value, called  $T_{c}$, where it presents a sharp drop.
Within the simple mean field approximation, this temperature, $T_{c}$, represents the passage from a superfluid regime  to a non-superfluid one.
This can be simply verified, by comparing the values given in Fig.\ref{CVPNM}, from the one we can extract from Fig.\ref{gapPNMT} (a) where the pairing gap goes to zero.

\section{Results}\label{Sect:results}

\subsection{Density}
In this section we discuss the results obtained by solving the HFB equations, Eq.\ref{HFBeq}, for the WS cells given in Tab.\ref{tabWS}.
As discussed already in article I, we do not show the results for the cell $^{982}$Ge, since the calculations, also at finite temperature, do not converge toward a stable configuration.
This problem of convergency could be due to the presence of a deformed minimum that is more energetically favorable~\cite{giaiGogny}.
If we consider the occupation probabilities of some given Hartree-Fock level at zero temperature, they are just 0 or 1. When we heat the system, this is no more true, since we introduce  in the equations  a Fermi-Dirac distribution that fragments the occupations of the levels (to some extent, similarly to the fragmentation introduced by a  pairing interaction).
As discussed recently in ref.~\cite{Margueron12}, this extra redistribution of the occupation of levels can favor the appearance of pairing correlations.
We report in Appendix B, the example of $^{176}$Sn; a double closed shell nucleus at T=0,  that becomes superfluid in a temperature range of $[0.05,0.63]$ MeV.

\begin{figure}
\begin{center}
\includegraphics[clip=,angle=-90,width=0.45\textwidth]{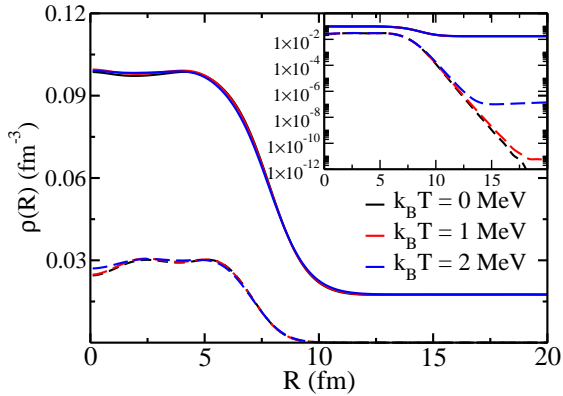}
\end{center}
\caption{
(Colors online) For the cell $^{1800}$Sn, the neutron (solid line) and proton (dashed line) density profile at three different temperatures $k_{B}T=0,1,2$ MeV. The calculations are done using the $V_{\text{low-k}}$ pairing interaction. In the inset we show the same result, but in semi-logarithmic scale. }
\label{denssn1800}
\end{figure}

In our calculations we work in the temperature interval [0-2] MeV, using a sampling step of $k_{B}\text{d}T=0.01$ MeV.
As an example, we show in Fig.\ref{denssn1800} the variation of the density as a function of the temperature for the cell $^{1800}$Sn calculated using the $V_{\text{low-k}}$ pairing interaction.
We observe that concerning the neutron, we do not have any remarkable variations going from $k_{B}T=0$ MeV to $k_{B}T=2$ MeV. Since this variation of the occupations due to the temperature happens around the Fermi energy, placed in the continuum, without any significant change on the neutron density.

\begin{figure*}
\begin{center}
\includegraphics[clip=,angle=-90,width=0.45\textwidth]{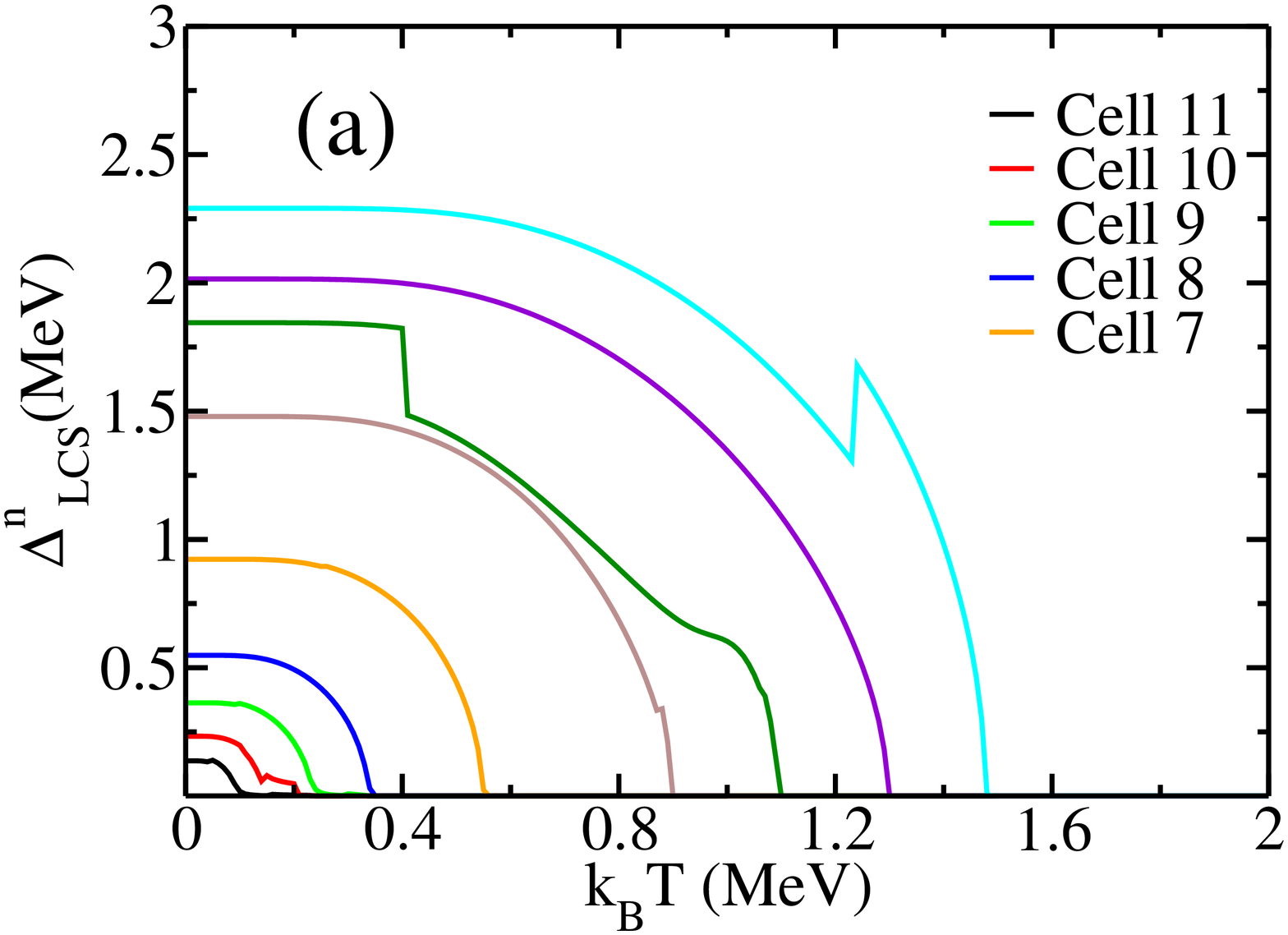}
\includegraphics[clip=,angle=-90,width=0.45\textwidth]{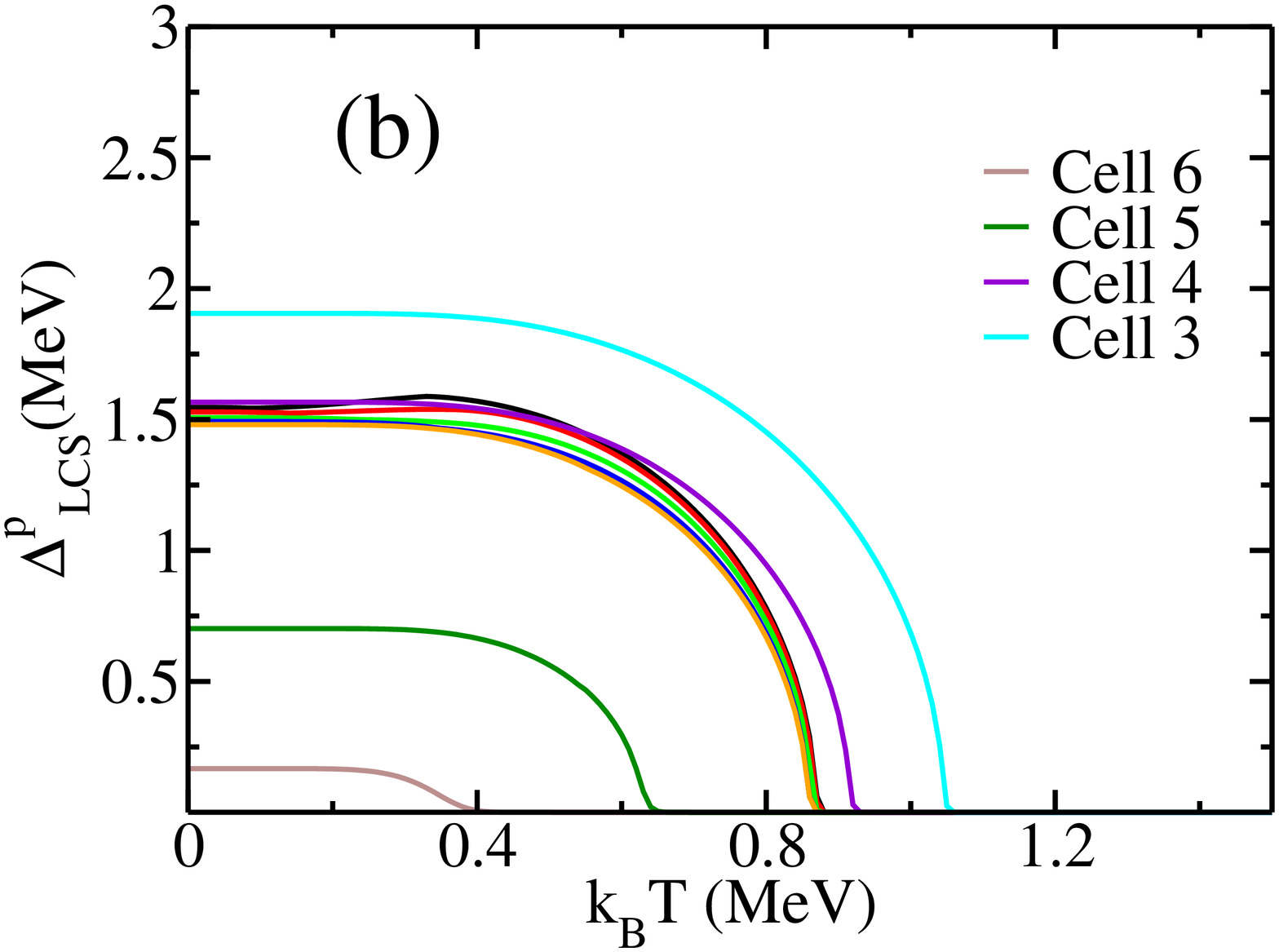}\\
\includegraphics[clip=,angle=-90,width=0.45\textwidth]{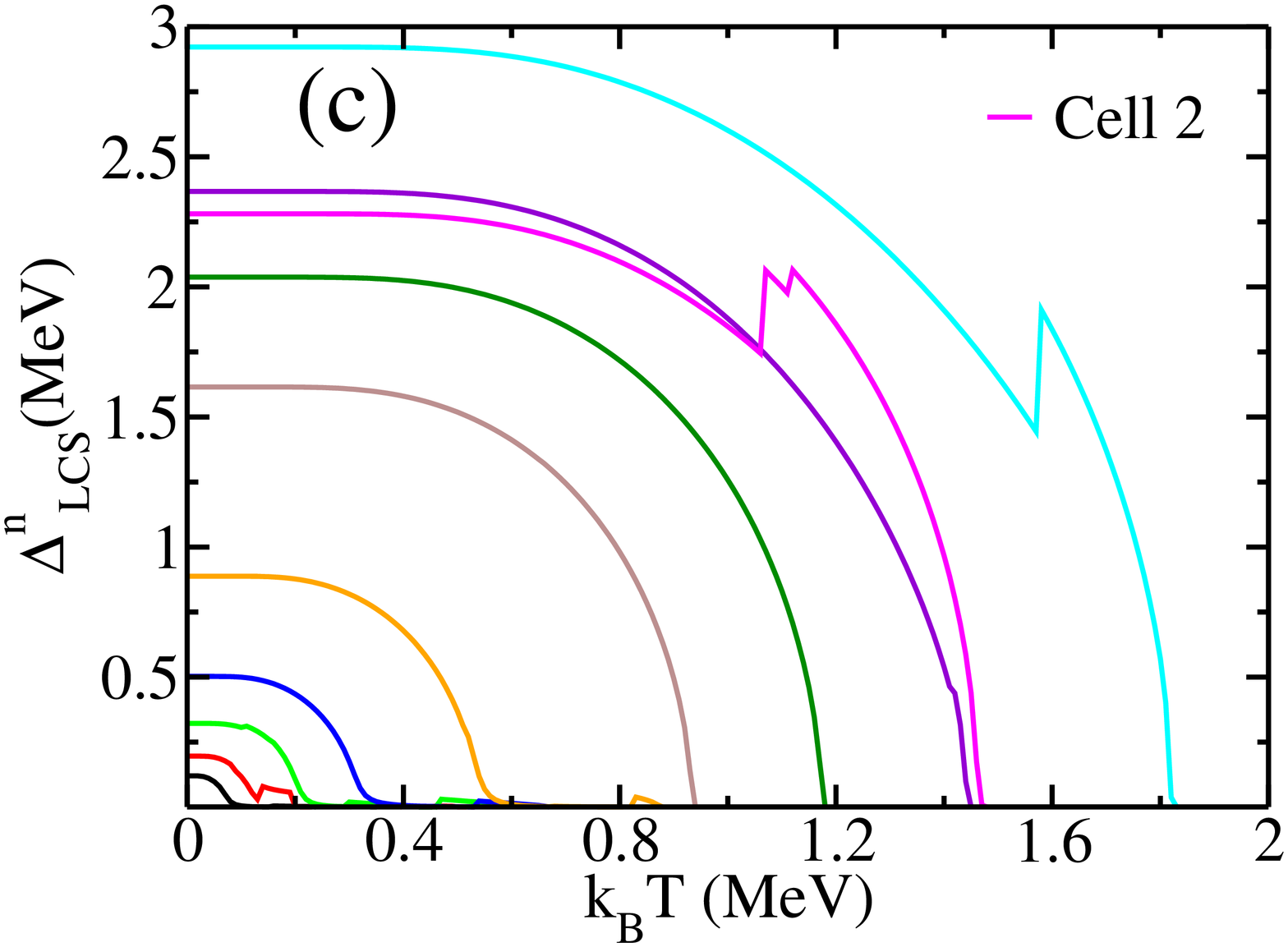}
\includegraphics[clip=,angle=-90,width=0.45\textwidth]{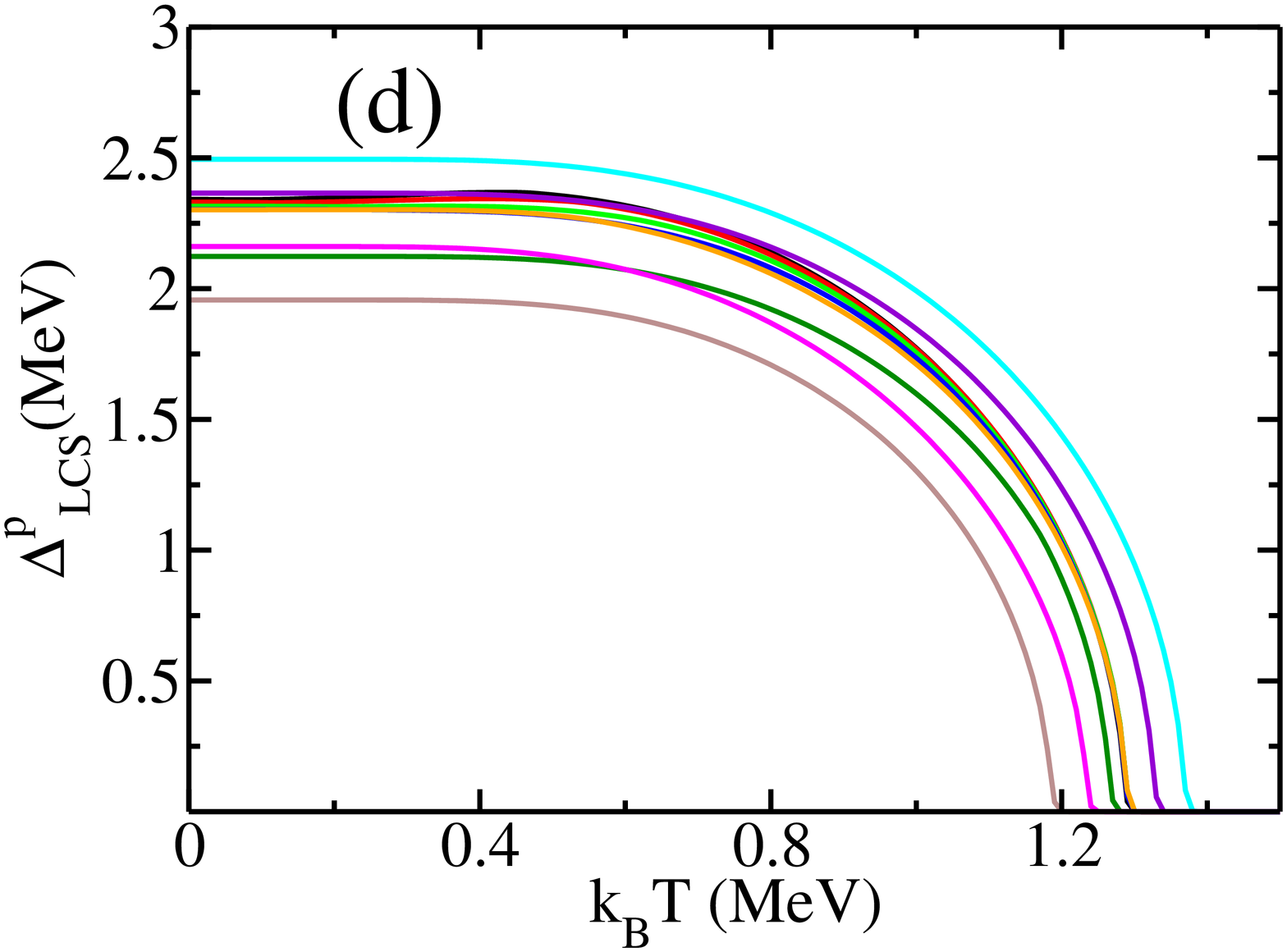}\\
\includegraphics[clip=,angle=-90,width=0.45\textwidth]{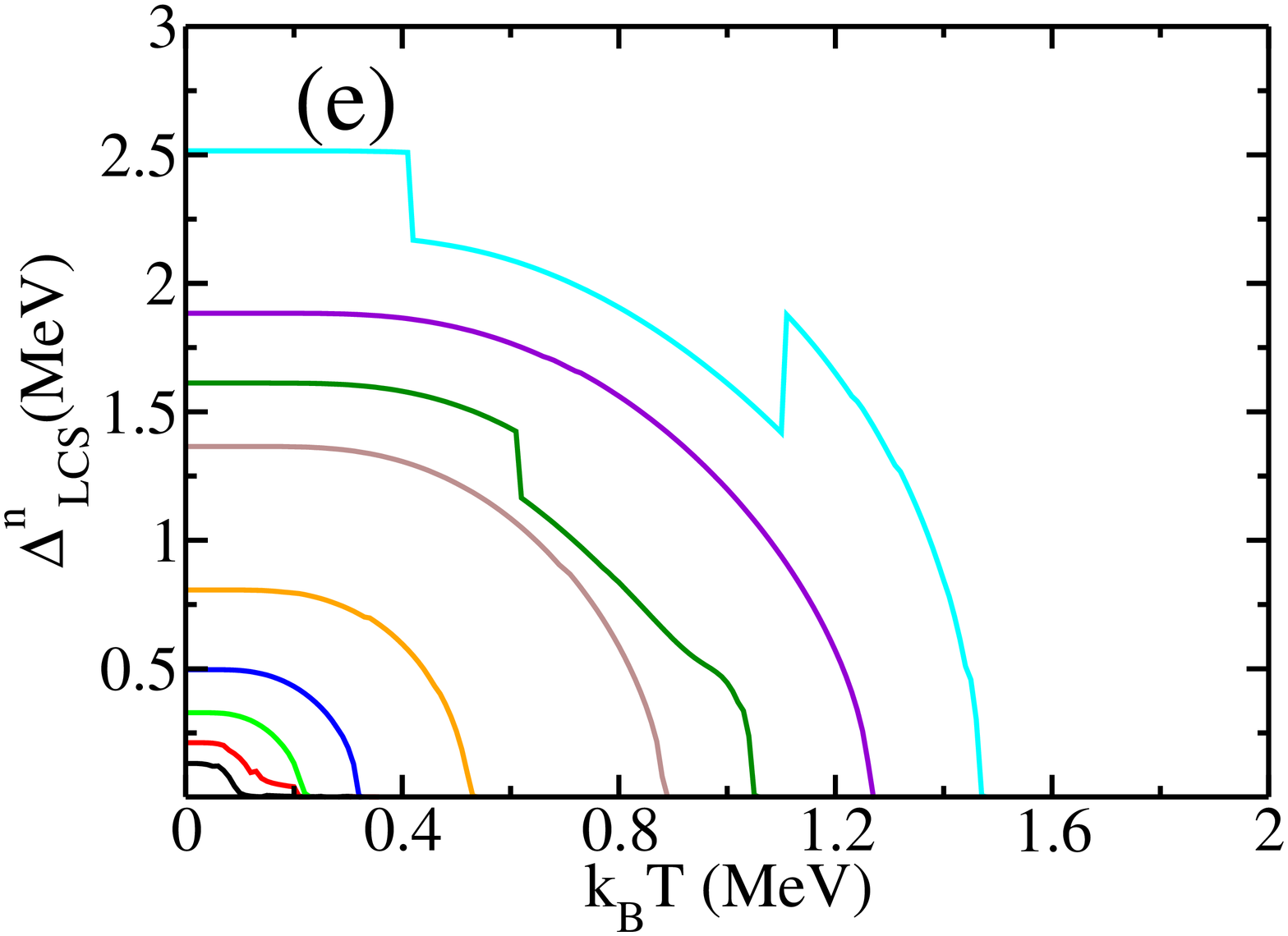}
\includegraphics[clip=,angle=-90,width=0.45\textwidth]{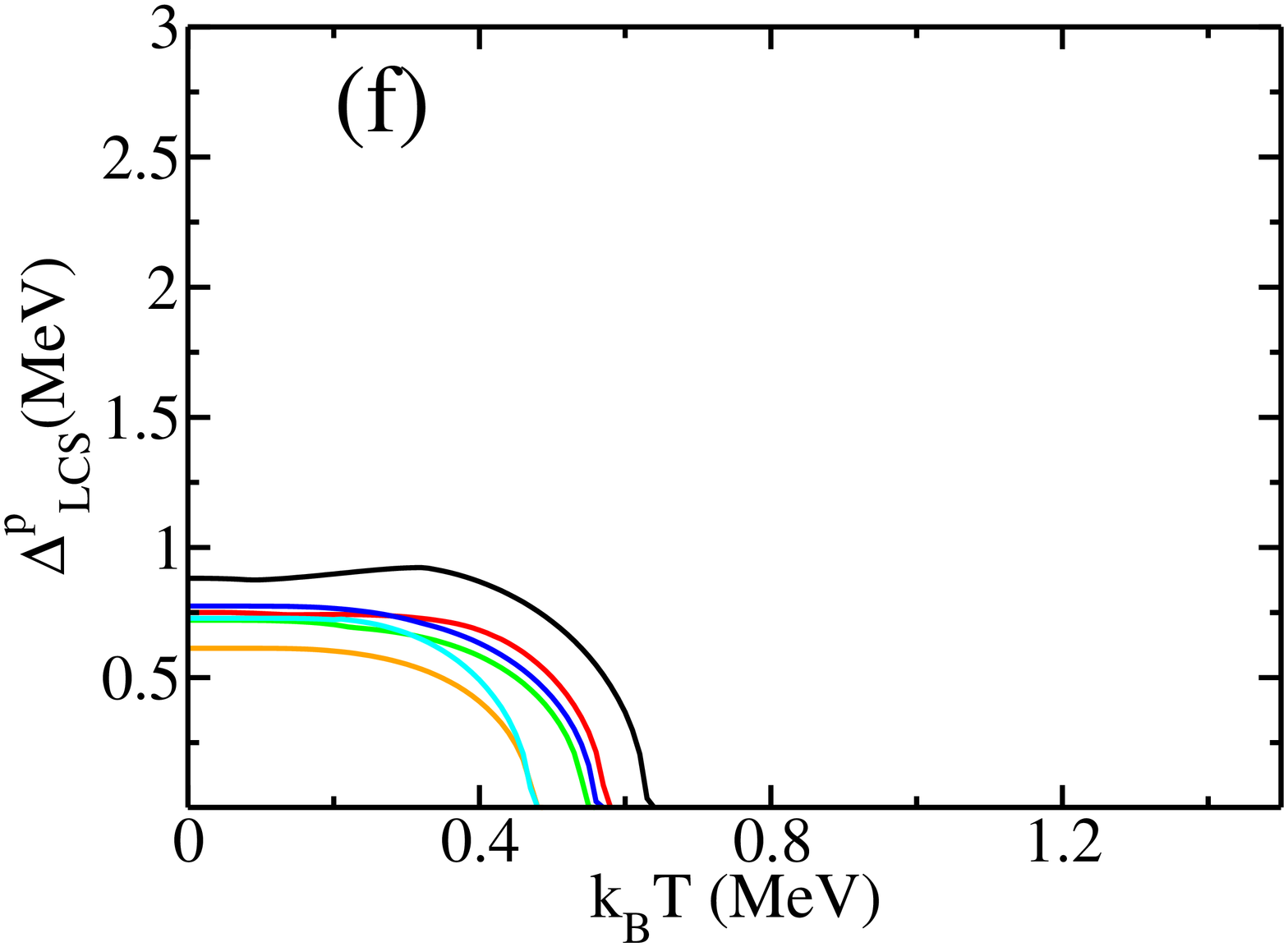}
\end{center}
\caption{
(Colors online) From the top to the bottom, neutron (left) and proton (right) LCS gaps, $\Delta^{q}_{LCS}$, as a function of the temperature and for the different cells given in Tab.\ref{tabWS}. In panel (a)-(b) we show the results for the  $V_{low-k}$ interaction, in the panels (c)-(d) the results for the Gogny D1 interaction and finally in panels (e)-(f) the results for the contact interaction DDDI.
For the different panels, we adopt the same color code. See text for details.}
\label{gapDLCS}
\end{figure*}

\noindent On the contrary, the proton density is more sensible to the effect of the finite temperature.  At $k_{B}T=0$ MeV the proton density decreases exponentially, since the protons occupy bound states, but with the increase of the temperature the proton can occupy scattering states. 
At $k_{B}T=1$ MeV the density of the proton gas is about $\rho^{gas}_{p}\approx 10^{-11}\text{ fm}^{-3}$ and it reaches $\rho^{gas}_{p}\approx 10^{-7}\text{ fm}^{-3}$ at $k_{B}T=2$ MeV.
We can calculate the number of protons in the gas region using the formula 

\begin{equation}\label{diff}
N_{p}^{gas}=\int_{R_{0}}^{R_{WS}} 4\pi r^{2}dr \left[\rho^{T}_{p}(r)-\rho_{p}^{T=0}(r)\right],
\end{equation}

\noindent where $R_{0}$ is the starting point of the integrations and is chosen to be $R_{0}\approx10$fm, $\rho^{T}_{p}(r)$ is the proton density at finite temperature and  $\rho_{p}^{T=0}(r)$ is the proton density at zero temperature.
Since we are interested only to the protons that occupy scattering states, in Eq.\ref{diff} we subtract the tail of the proton density at zero temperature. In such a way the result is not very sensitive to the choice of $R_{0}$.
We thus obtain for $k_{B}T=1$MeV the value of $N_{p}^{out}\approx0.008$, while for  $k_{B}T=2$MeV  we have $N_{p}^{out}\approx0.05$ protons.
A detailed analysis about the appearance of a proton gas for finite nuclei at finite temperature has been already presented  by P. Bonche \emph{et al.} \cite{Bonche84}, showing that the vapor is present because we impose our system to be confined in a box, otherwise the Coulomb repulsion should provoke the evaporation of this gas. Thus to have  box-independent results, Bonche et collaborators suggested a vapor subtraction procedure.

Since we work within the WS approximation, the possibility of the presence of a proton gas is admitted, since the confinement is imposed by the presence of the other surrounding cells, although in such case the WS approximation itself could be less justified. For the temperature regime in which we work, this proton gas is very weak and we will not apply any specific procedure , but we remark that this phenomenon should be better investigated in the future.

\subsection{Pairing gaps}

\begin{figure}
\begin{center}
\includegraphics[clip=,angle=-90,width=0.45\textwidth]{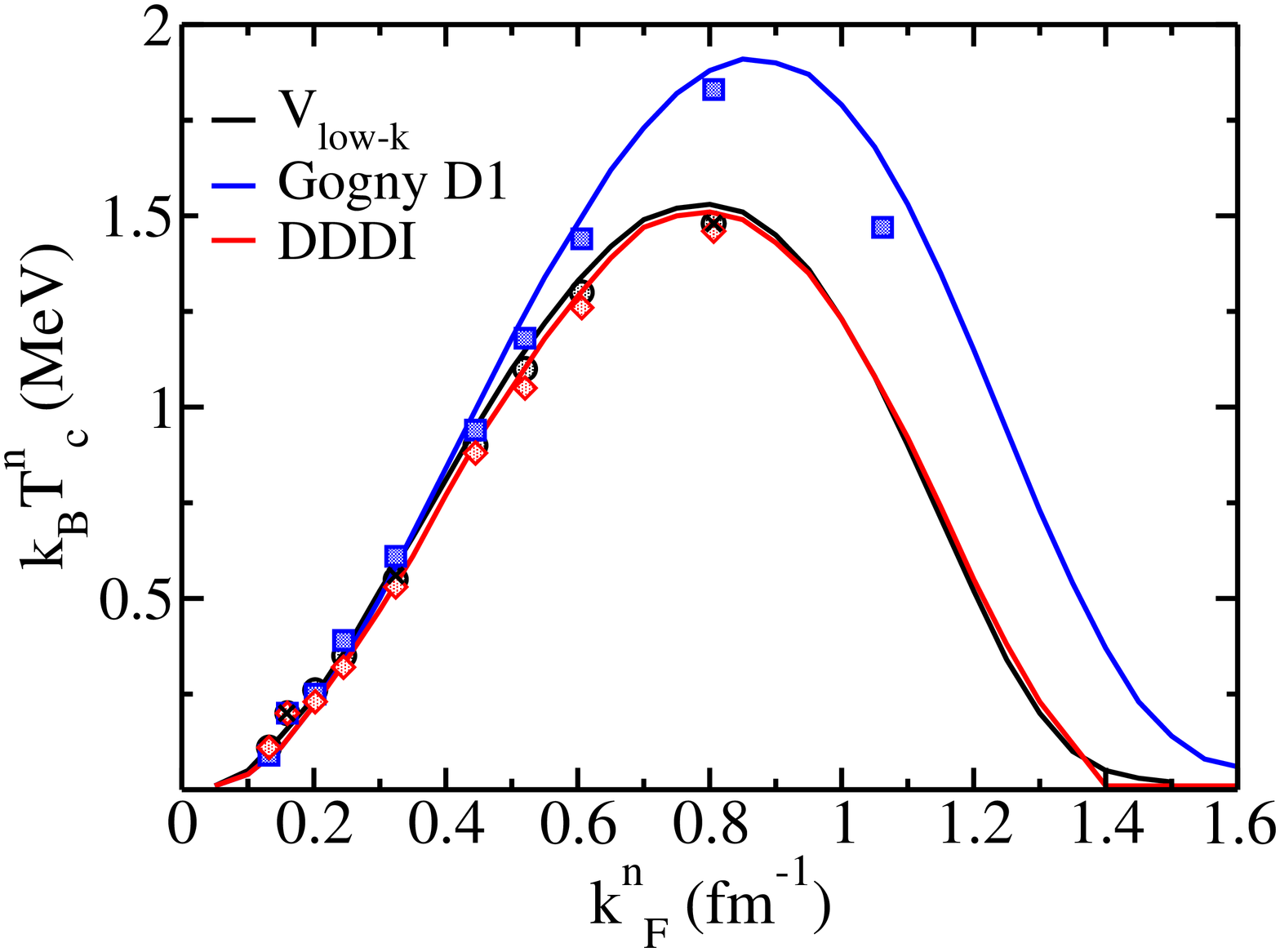}
\end{center}
\caption{
(Colors online) Critical temperature for neutrons $k_{B}T^{n}_{c}$ calculated using the  three different pairing interactions discussed in the article. The solid lines represent the pure neutron matter result calculated using Eq.\ref{gapeqINM}, while the points represent the WS calculations calculated using the BCE. On the same figure the crosses represents the calculations for three WS cells using $V_{\text{low-k}}$ pairing interaction and BCO conditions. See text for details. }
\label{Tcrit}
\end{figure}

We discuss here the results we obtained from our HFB calculations at finite temperature, with special attention to the pairing gap.
In Fig.\ref{gapDLCS} we show the neutron and proton pairing gaps $\Delta_{LCS}^{q}$ using the three different pairing interactions: $V_{low-k}$ (panel (a)-(b)), Gogny D1 (panel (c)-(d)) and the contact interaction (panel (e)-(f)). 
In all the figures of this article we present the results from $^{180}$Zr to $^{1800}$Sn for all the pairing interactions plus  the results of $^{1500}$Zr for the Gogny D1, since only with this interaction the results converge to a stable solution (see discussion in article I).

We observe that as a common feature, the value of the LCS gap decreases as we increase the value of the temperature and it finally disappears when reaching a specific value of the temperature, called $T^{q}_{c}$. A similar behavior has been already observed for the PNM system in Fig.\ref{gapPNMT} (a).
Compared to the case of pure neutron matter, we observe that the  $\Delta_{LCS}^{n}$  is not a smooth function of the temperature $k_{B}T$, but it presents some \emph{jumps}. This is a consequence of Eq.\ref{gapLCS} that selects the pairing gap of the single-particle canonical state with the energy closest to the Fermi energy.
As a result, it can happens that we do not always select the same level for different values of the temperature, since there could be a change of the single particle structure due to the temperature  and the \emph{jumps} correspond to the selection of a different state in Eq.\ref{gapLCS}.
Another definition of pairing gap averaged on more levels around the Fermi energy would avoid this problem ($i.e.$ see Eq.\ref{gapuv}), but it would have as inconvenient to mix the superfluid properties of the bound states, with the one of the external gas.
For the case of protons, the two definitions of pairing gap give very similar results, since they are bound.

  A more detailed analysis on the properties of the gap LCS   will be the subject of a forthcoming article \cite{PastoreSchuck}, but we can anticipate here that Eq.\ref{gapDLCS} is a good filter in the case of WS cells to study the superfluid properties of the external gas, where we have usually the majority of matter in each  cell.
  
Comparing the panel (a) with  the panel (e) of Fig.~\ref{gapDLCS} we observe that the zero range force reproduces quite nicely the global trend and the absolute values of the neutron gaps as a function of the temperature. with a discrepancy of less than $\approx$10\% on the values of the gaps obtained with the two forces. That is not the case for the protons, where there is almost a factor of 2 difference in the gaps, but both interactions predict a superfluid solution for protons (on the contrary of the DDDI adopted in article I).

 The origin of the difference comes from the fitting protocol we adopted to determine our density dependent interaction (Eq.\ref{pairing_int_contact}), since we used only infinite matter properties ($i.e.$ the value of the pairing gap in SNM and PNM) without taking into account  any kind of shell effects ($i.e.$ including some finite nuclei gaps into our $\chi^{2}$ minimization to determine the parameters $\eta_{i=1,2},\alpha_{i=1,2}$).
We observe that the analytic  formula obtained from BCS theory~\cite{Grill_2011,bcsbook}

\begin{equation}\label{empiric}
k_{B}T^{q}_{c}\approx0.56\cdot \Delta^{q}_{k_{B}T=0}\,,
\end{equation}

\noindent  used to derive the critical temperature  from the value of the gap at zero temperature, $\Delta^{q}_{k_{B}T=0}$,  is compatible with our findings. 

To understand the role of the bound nucleus in the center of the WS cell, in Fig.\ref{Tcrit} we compare the value of the critical density $T^{n}_{c}$ at which the gap $\Delta_{LCS}^{n}$ for neutron disappears, in both  PNM  ($i.e$ solving Eq.\ref{gapeqINM}) and WS cell.
We associate at each WS cell a neutron Fermi momentum, $k_{F}^{n}$, according to the value of the density of the external neutron gas, $\rho_{n}^{b}$, as given in Tab.\ref{tabWS}.
We observe that similarly to the zero temperature case (see article I), the presence of the cluster reduces the neutron superfluidity for the high density cells. It can be seen from a smaller critical temperature for the WS cell compared to the PNM case. 
In Fig.\ref{Tcrit} it is not easy to clearly see such difference, but it will appear more clear when we will compare  the specific heat of the two systems  as  done in Fig.\ref{Tcrit2} and Fig.\ref{Tcrit3}, we thus refer to the discussion in the next section. 
The Gogny D1 interaction presents the same qualitative  behavior with the difference that usually the critical temperature, $T_{c}$, appears at higher temperatures due to the stronger pairing strength compared to the $V_{\text{low-k}}$ interaction (or the DDDI). This is what one can expect by simply using Eq.\ref{empiric} for an estimation.  We refer to  Appendix A for a more detailed discussion.

As a last remark, we observe that the  DDDI interaction given in Eq.\ref{pairing_int_contact} gives a very similar value  of critical temperatures for neutrons as the $V_{\text{low-k}}$ pairing interaction, as expected from the results of Fig.\ref{gapDLCS}, at most with a difference of $\approx$30 KeV.
Such small difference can not be clearly seen on this picture, but it becomes much more clear in Fig.\ref{CVT}(a)(b), where we present the specific heat.
In some cells and for few values of temperature $T$, our numerical code was not able to converge the HFB equations using the DDDI pairing interaction. The code oscillates among two possible solution that are relatively close in energy ( $\Delta E \approx$1 MeV). We tried changing the damping parameter \cite{karim} without any remarkable improvement. 
In this case we do not include these points in our calculations of the entropy and of the specific heat.

We also tested the dependence of our results on the choice of the boundary conditions.
As shown already in article I and also in ref.\cite{Avogadro08}, the BCE and BCO boundary conditions give  similar results concerning gaps at $k_{B}T$=0 MeV.
We checked that is true also at finite temperature. We performed calculations in 3 different cells: $^{200}$Zr,$^{500}$Zr and $^{1800}$Sn using the $V_{low-k}$ paring interaction and BCO conditions. 
In Fig.\ref{Tcrit} we report the result of such calculations for the case of critical temperature. The differences among the critical temperatures obtained with the  BCE and BCO conditions is at most $\approx$10 KeV in the cell $^{500}$Zr, showing that the results are robust and do not depend on the particular choice of boundary conditions.
This is in contrast with the results shown in ref.\cite{Baldo_2006}, the main difference among the two calculation is that Baldo and collaborators minimized a new set of WS cells using the different boundary conditions, while in this work we limit to the WS cells calculated by Negele and Vautherin~\cite{Negele_1973}.
\noindent We also refer the reader to Appendix C for further discussions on the \emph{finite-size} effects and boundary conditions.

\subsection{Specific heat}

Exactly as we did for the PNM system, we now examine the entropy and the specific heat for the WS cell.
The expressions for these quantities have been already given in Eq.\ref{EqentropyPNM} and Eq.\ref{spec_heatPNM}, for the infinite system and it straightforward to adapt them for the calculations in the WS cells.

\begin{figure*}
\begin{center}
\includegraphics[clip=,angle=-90,width=0.45\textwidth]{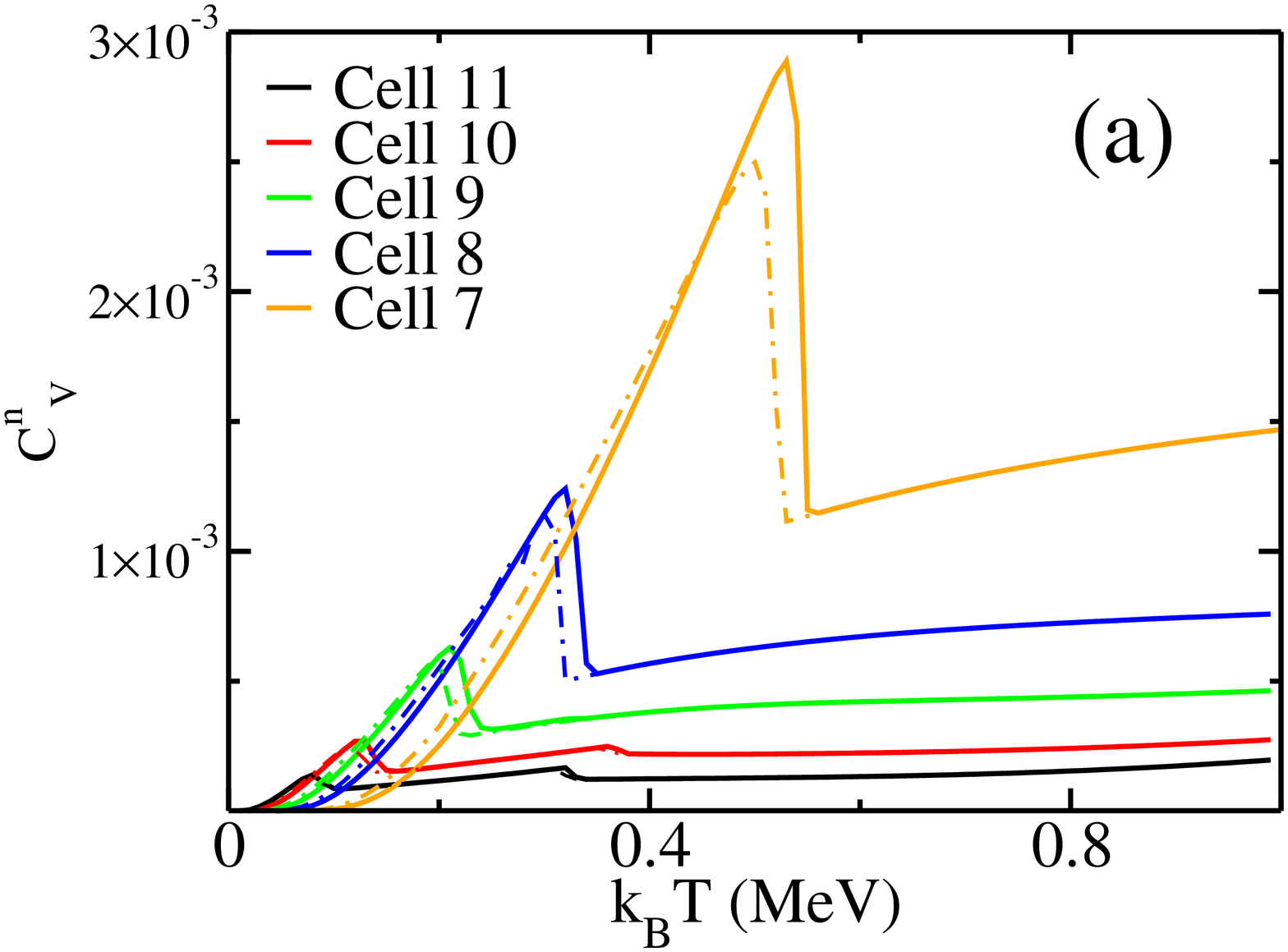}
\includegraphics[clip=,angle=-90,width=0.45\textwidth]{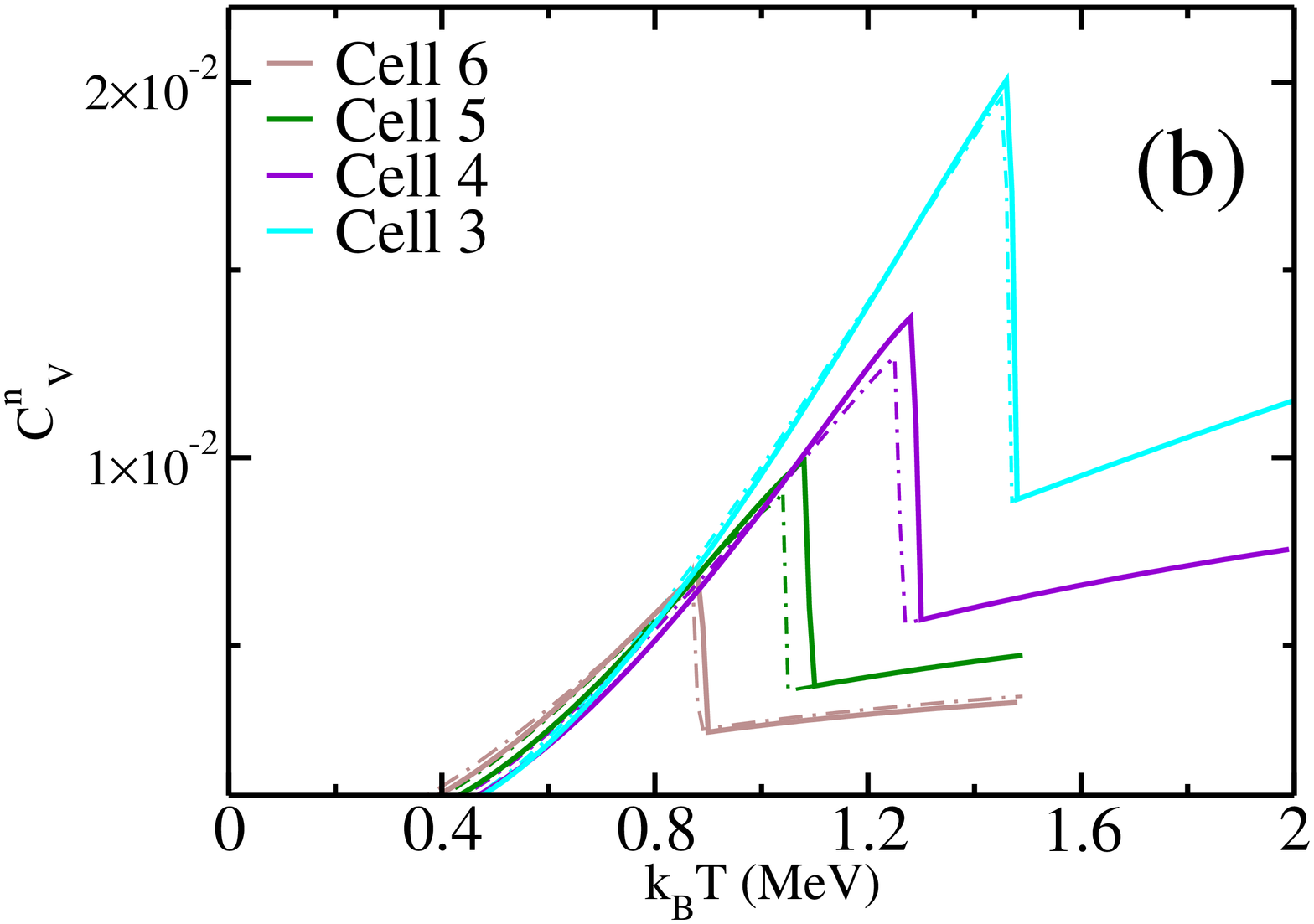}
\includegraphics[clip=,angle=-90,width=0.45\textwidth]{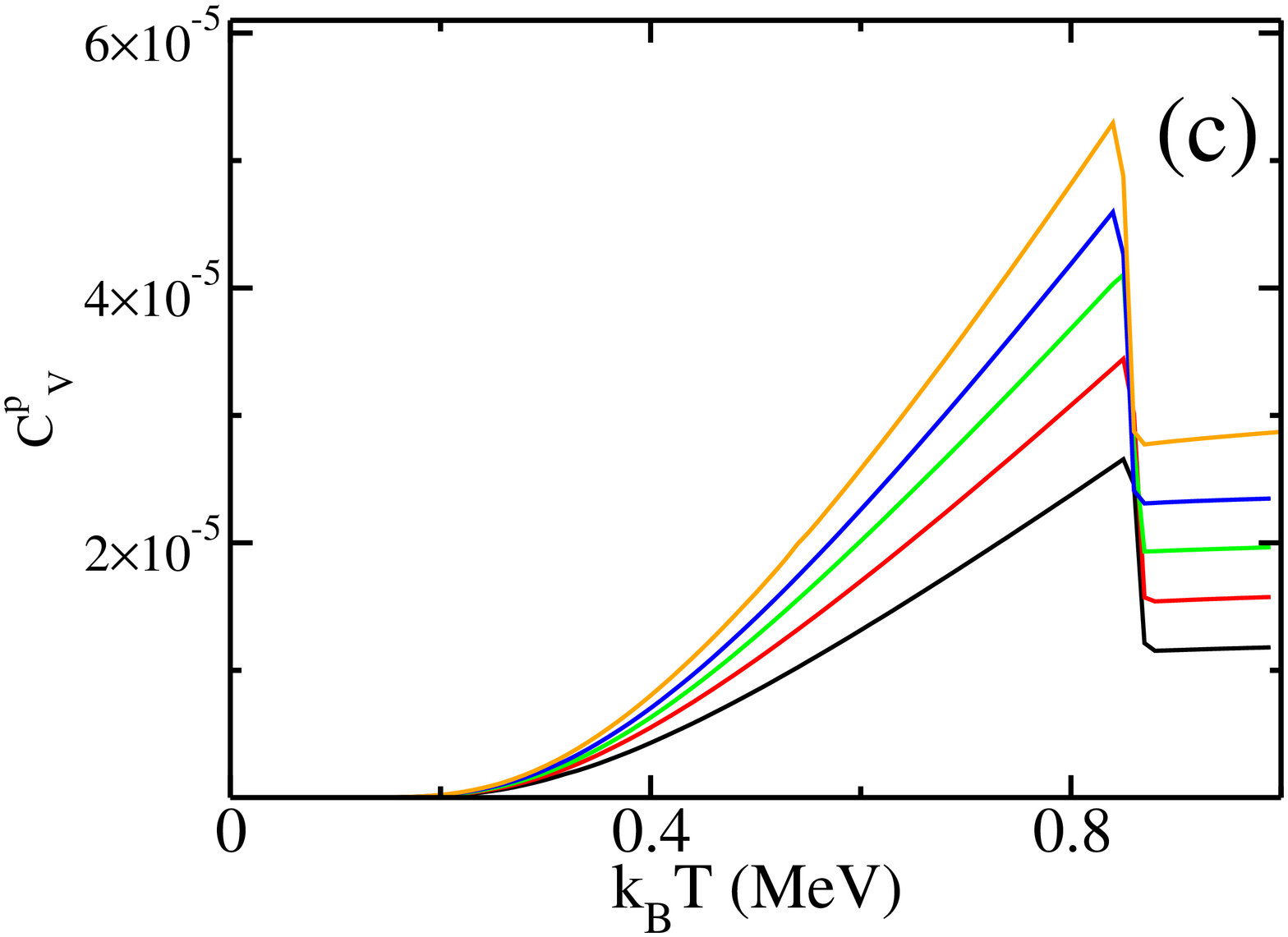}
\includegraphics[clip=,angle=-90,width=0.45\textwidth]{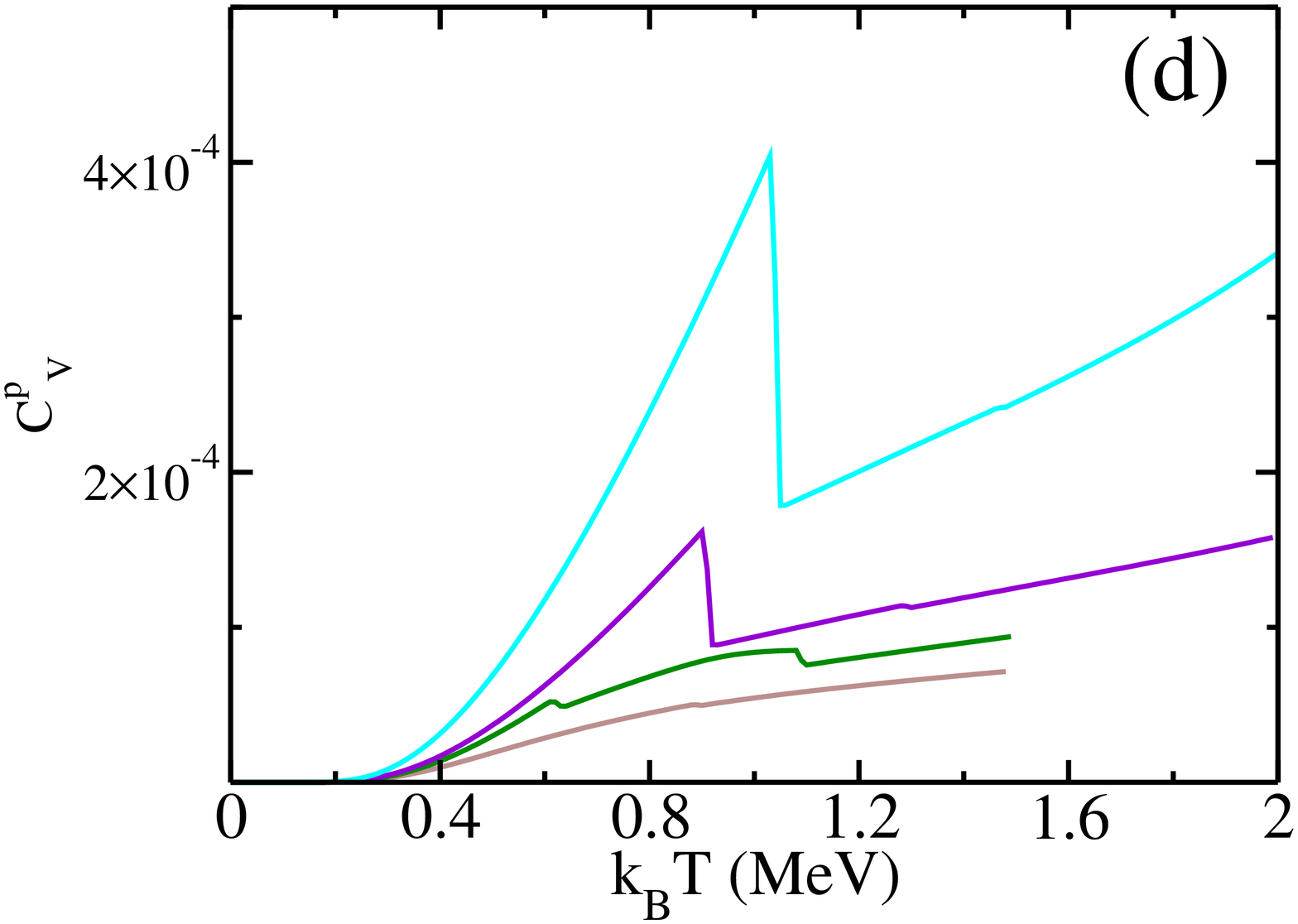}
\end{center}
\caption{
(Colors online) On the top row (a-b) the neutron specific heat $C_{v}^{n}$  as a function of the temperature for the different cells of Tab.\ref{tabWS} obtained with $V_{low-k}$ pairing interaction, on the bottom (c-d) the specific heat $C_{v}^{p}$ for protons . In panel (a-b) we also show, as dashed-dotted line, the neutron specific heat $C_{v}^{n}$ obtained using the DDDI pairing interaction given in Eq.\ref{pairing_int_contact}. See text for details.
}
\label{CVT}
\end{figure*}

\noindent In Fig.\ref{CVT} we show the specific heat of protons and neutrons calculated using the $V_{\text{low-k}}$ interaction. In each cell the proton specific heat is usually two orders of magnitude smaller then the neutron specific heat.
As already observed in Fig.\ref{CVPNM}, the sudden variations in the $C_{V}^{q}$ are associated with a phase transition from superfluid to non-superfluid system. This can easily understood comparing the position of the peaks with the values of the critical temperature $T_{c}^{q}$ given in Fig.\ref{Tcrit}.
In the same figures in panels (a) (b) we also show the specific heat $C_{V}^{n}$ calculated using the contact interaction given in Eq.\ref{pairing_int_contact}. We observe that the curves follow quite closely the one obtained using the $V_{\text{low-k}}$ interaction, apart from a small difference for the high density cells. This is a direct consequence of the differences we observed in the Figs.\ref{gapDLCS} and \ref{Tcrit}, since the DDDI interaction is able to mimic the superfluid properties of the $V_{\text{low-k}}$ only within a certain error.

\begin{figure}
\begin{center}
\includegraphics[clip=,angle=-90,width=0.45\textwidth]{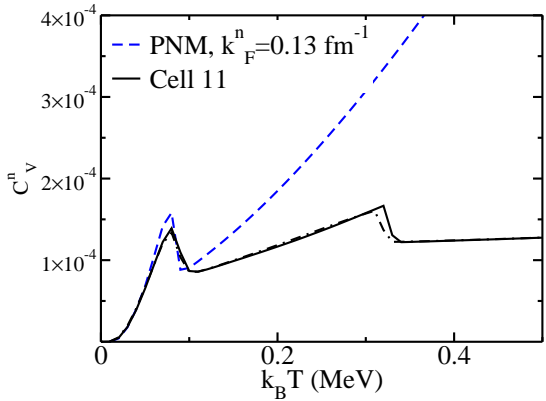}
\end{center}
\caption{
(Colors online) The neutron specific heat for $^{180}$Zr calculated with the $V_{\text{low-k}}$ pairing interaction (solid line), and with the DDDI pairing interaction (dashed-dotted line). In the same figure we show the corresponding PNM calculations (dashed line) at $k_{F}^{n}=0.12$ fm$^{-1}$ (see Tab.\ref{tabWS}).}
\label{Tcrit2}
\end{figure}

\begin{figure}
\begin{center}
\includegraphics[clip=,angle=-90,width=0.45\textwidth]{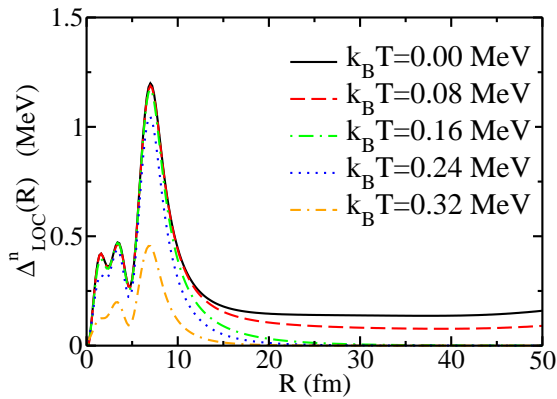}
\end{center}
\caption{
(Colors online)Local pairing field of neutrons, $\Delta_{LOC}^{n}(r)$, for the WS cell $^{180}$Zr at different temperatures calculated using the $V_{\text{low-k}}$ pairing interaction. See text for details.}
\label{Dloczr180}
\end{figure}

\begin{figure}
\begin{center}
\includegraphics[clip=,angle=-90,width=0.45\textwidth]{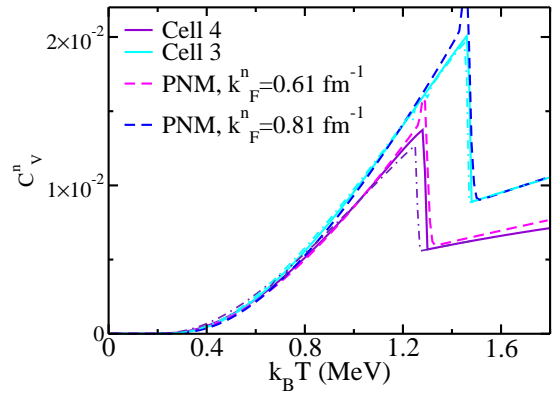}
\end{center}
\caption{
(Colors online) The neutron specific heat for the two cells $^{1350-1800}$Sn (solid line) and the corresponding PNM calculations (dashed line)  at $k_{F}^{n}=0.6,0.8$ fm$^{-1}$ (see Tab.\ref{tabWS}) calculated using the $V_{\text{low-k}}$ pairing interaction. In the same figure we show with the same color code, the results of WS cells calculated using the DDDI interaction (dashed-dotted line). See text for details.}
\label{Tcrit3}
\end{figure}

 In Fig.\ref{CVT} (a), we observe that the low-density cells present two peaks. 
To better understand this phenomenon in Fig.\ref{Tcrit2}, we compare the neutron specific heat $C^{n}_{V}$ for the cell $^{180}$Zr and the specific heat of PNM calculated at the neutron Fermi momentum $k_{F}^{n}=0.13$ fm$^{-1}$.  See Tab.\ref{tabWS}.

The two systems show a similar behavior up to the critical temperature, $k_{B}T^{n}_{c}\approx0.1$ MeV, where we have the first phase transition. For PNM the peak is located at $k_{B}T^{n}_{c}=0.08$ MeV and it corresponds to the phase transition from superfluid to non-superfluid system (as discussed in Fig.\ref{CVPNM}), then the $C^{n}_{V}$ monotonically increases. 
On the contrary for the $^{180}$Zr cell we have a first peak at $k_{B}T^{n,1}_{c}=0.09$ MeV and a second one at $k_{B}T^{n,2}_{c}=0.32$ MeV.
A similar result  has been already discussed in ref.\cite{Grill_2011}, adopting a simple contact force with no isovector dependence ($i.e.$ putting $f_{2}^{q}\equiv0$ in Eq.\ref{pairing_int_contact} ).

We can deduce that the first transition refers to the disappearance  of the neutron superfluidity for the external neutron gas, $k_{B}T^{n,1}_{c}$, while the bound neutrons in the cluster are still superfluid. The second peak at $k_{B}T^{n,2}_{c}$ corresponds to the transition of the bound neutrons from superfluid phase to non superfluid phase.
To confirm this hypothesis, we analyze the local pairing field, $\Delta^{n}_{\text{LOC}}(R)$,  of neutrons for $^{180}$Zr. We refer to article I for the exact definition of this quantity (see also ref.\cite{Pastore_2008}).

In Fig.\ref{Dloczr180} we show  $\Delta_{LOC}^{n}(R)$ for the WS cell $^{180}$Zr at different temperatures calculated using the $V_{\text{low-k}}$ pairing interaction.
We clearly see that at the first critical temperature, $k_{B}T^{n,1}_{c}$, the external neutron gas becomes non-superfluid, since the pairing field goes to zero, but it remains non-zero inside the nucleus without significative changes.
At the second critical temperature, $k_{B}T^{n,2}_{c}$, the superfluidity disappears also from the nucleus, and the entire cell becomes non-superfluid.
We conclude that the value of the critical temperature of neutrons given in Fig.\ref{Tcrit} corresponds to the phase transition of the gas, since as we discussed previously in the text, we used as a probe to study the  superfluid properties the gap LCS, $\Delta^{n}_{\text{LCS}}$, given in Eq.\ref{gapLCS}. As anticipated in previous section, this is a feature of the gap LCS valid for the WS cell system.

According to Fig.\ref{CVT} (a) we have 2 phase transitions in the cells: $^{180}$Zr, $^{200}$Zr and $^{250}$Zr. As expected we find the same behavior, with the two phase transitions when using the DDDI interaction, confirming that this is a general property of the system.
In Fig.\ref{Tcrit3} we repeat the comparison already done in Fig.\ref{Tcrit2}, but for the cells $^{1350}$Sn and $^{1800}$Sn.
Again the DDDI and the $V_{\text{low-k}}$ interactions give very similar results.
We do not observe any major qualitative difference when adopting the Gogny D1 pairing interaction, we thus refer the reader to Appendix A for a more  detailed discussion.

We now analyze the proton specific heat, $C_{V}^{p}$, calculated using the  $V_{\text{low-k}}$ pairing interaction. We observe that for the cells from 11 to 7 we have only one transition occurring at  temperature $k_{B}T_{c}\approx0.8$ MeV, see Fig.\ref{CVT} (c), that are usually at higher temperature than the corresponding phase transitions for neutrons.
More interesting is the behavior of proton specific heat for the last three cells. 
For example in $^{1100}$Sn  we observe a first peak at $k_{B}T_{c}^{p,1}=0.62$ MeV, corresponding to the passage superfluid/non-superfluid regime for protons ($i.e.$ zero proton pairing energy). We encounter a second peak at $T_{c}^{p,2}=1.08$ MeV this is the same value of the temperature of the neutron phase transition from the superfluid to non-superfluid phase $k_{B}T_{c}^{n,2}=1.08$~MeV.
We conclude that the modification of the neutron entropy affects the proton entropy trough the isovector interaction among neutron and protons at the mean field level. Since the number of neutrons is much bigger then the number of protons, we do not observe the opposite phenomenon ($i.e.$ a change in the proton entropy affecting the neutron entropy).
The results for neutrons obtained using the DDDI interaction are in good agreement with the results employing the $V_{\text{low-k}}$ pairing interaction as seen in Fig.\ref{Tcrit2}-\ref{Tcrit3}. This is not the case for protons, where the two interaction gives quite different quantitative results.

\section{Conclusions} \label{Sect:conclusions}

We have performed extensive calculations of Wigner Seitz cell adopting three different pairing interactions: two pairing interactions with finite range, but with different  strengths, $i.e.$ $V_{\text{low-k}}$ and Gogny D1, and a density dependent contact one fitted to reproduce the infinite matter results given by the  $V_{\text{low-k}}$ interaction.
Since we used the same numerical code for all the forces, we have been able to make a strict comparison between them.
Comparing with the results obtained in pure neutron matter, we investigated the role of the strength and of the range of the pairing interaction on the specific heat of the inner crust.
As a result, we observed that the features of neutron superfluidity described by the $V_{\text{low-k}}$ interaction and our DDDI are essentially the same.
This is not the case for the protons, where the two interactions are in agreement only at the qualitative level.
We also investigated the effect of the strength of the pairing functional on the superfluid properties of the WS cells.
In particular we confirm, as in article I, that this effect is more  important for the high density cells, where pairing can also affect the configuration of the cell~\cite{Baldo_2006,Grill_2011}.
We observe that, independently of the adopted pairing interaction, the cells close to the outer crust shows two distinct phase transitions: the first one appears at low temperature and it concerns the external neutron gas, while the  second at higher temperature describe the transition of  the nucleus from superfluid to non-superfluid phase.
We think that the intermediate phase among the two critical temperatures, $i.e.$ a superfluid nucleus immersed in a sea of non-superfluid neutrons should be better investigated, since it could have important effects on the properties of the inner crust.
In fact in this case we have a significant deviation among the specific heat of the WS cell and the infinite system at the same density.
This investigation goes beyond the scope of the present article and we leave it for a future work.

We conclude by observing, that our results are obtained within the mean field approximation.
It has been observed by some authors  \cite{Gori_2004,Gori_2005,baroni,Soma}, that the inclusion of higher order diagrams, $i.e$ particle-vibration coupling,  can strongly affect the shell structure and the superfluid properties of the system.
Such kind of calculations are still not available for WS cells, mainly because of the computational time  requested.
According to our findings, a possible alternative  could be represented by performing calculations in the infinite systems of both collective excitations~\cite{PastoreSNM,PastorePNM,Margueron08b} and  pairing gaps \cite{Lombardo}. Such results could be then used to fit an effective zero range pairing interactions.
Finally the resulting contact force could be used to calculate the properties of the inner crust.
This method although the underlying approximations, could help giving a first quantitative estimate of such higher order effects in WS cells.

\section{Acknowledgments}

We are grateful to T. Lesinski and T. Duguet for useful discussions 
and for providing us with the separable representation of low-momentum realistic interactions.
We thank K. Bennaceur and D. Davesne for proof reading the manuscript and for interesting discussions.
We also thank S. Baroni, J. Margueron, S. Goriely, N. Chamel, J. M. Pearson and C. Losa  for useful and interesting discussions.
The author acknowledges the hospitality of the
Theory Group of the Institut de Physique Nucl\'eaire de
Lyon during the two post-doctoral years at the University
of Lyon.

\begin{appendix}
\section{Gogny D1}

\begin{figure*}
\begin{center}
\includegraphics[clip=,angle=-90,width=0.45\textwidth]{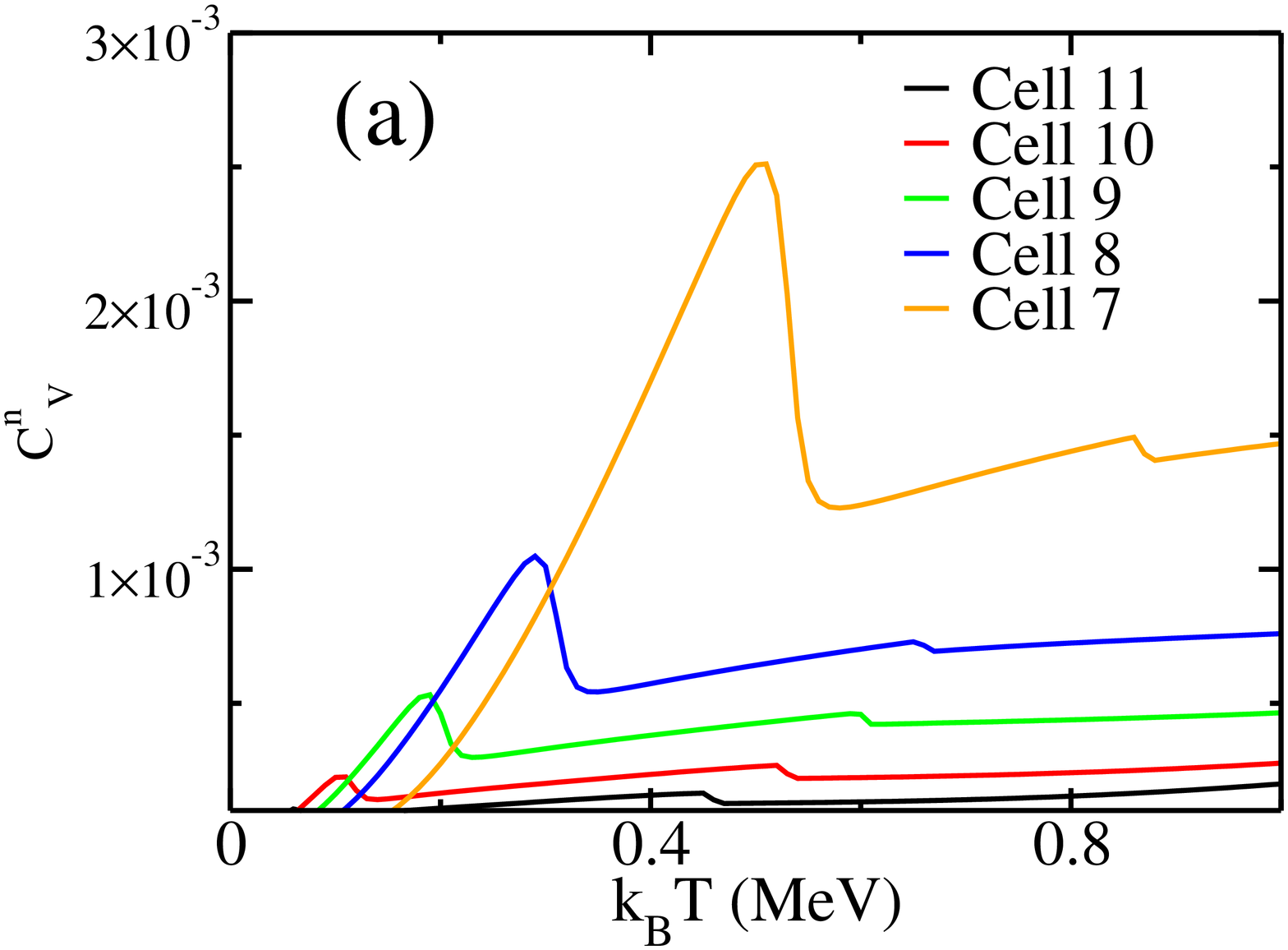}
\includegraphics[clip=,angle=-90,width=0.45\textwidth]{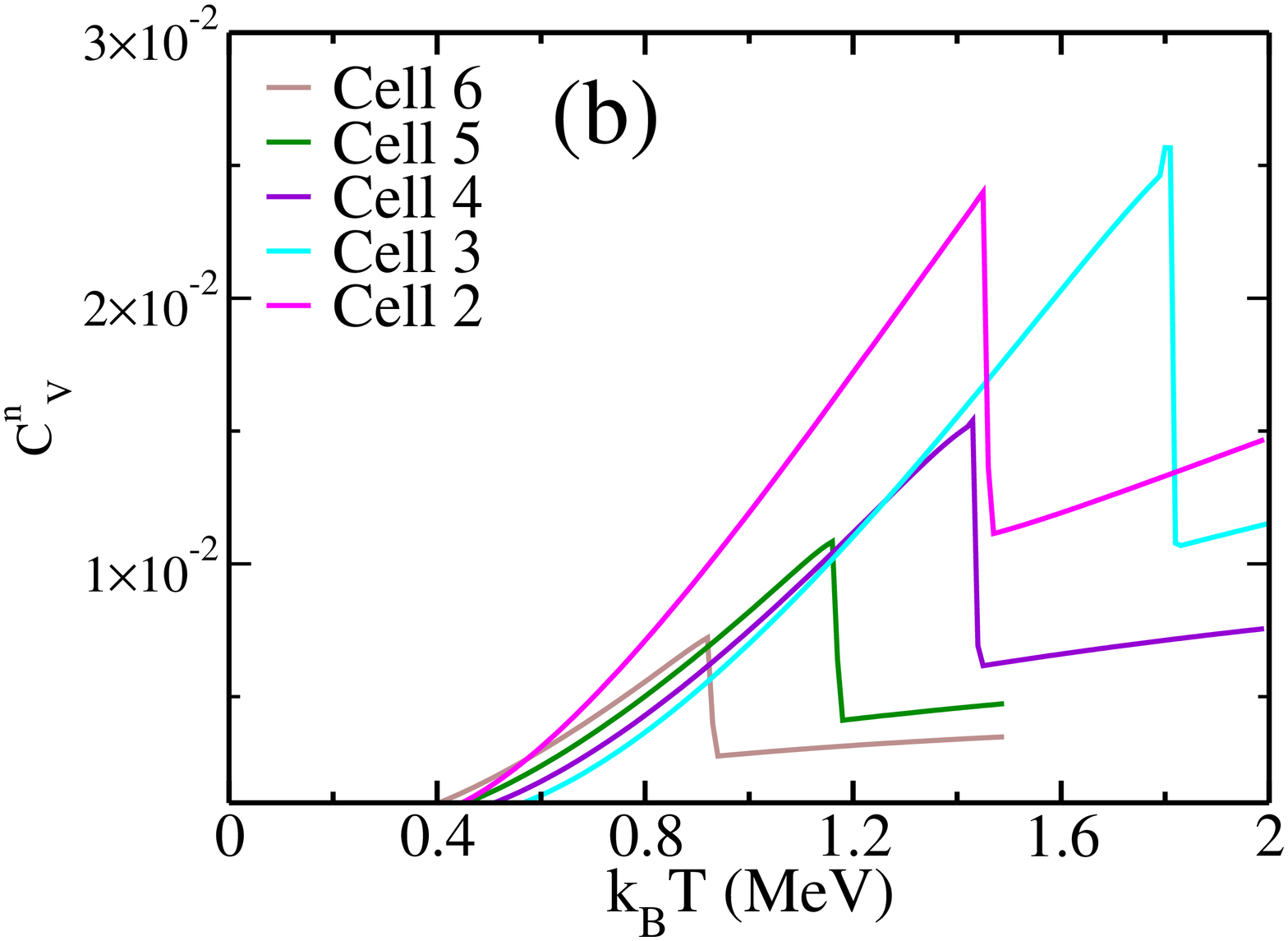}\\
\includegraphics[clip=,angle=-90,width=0.45\textwidth]{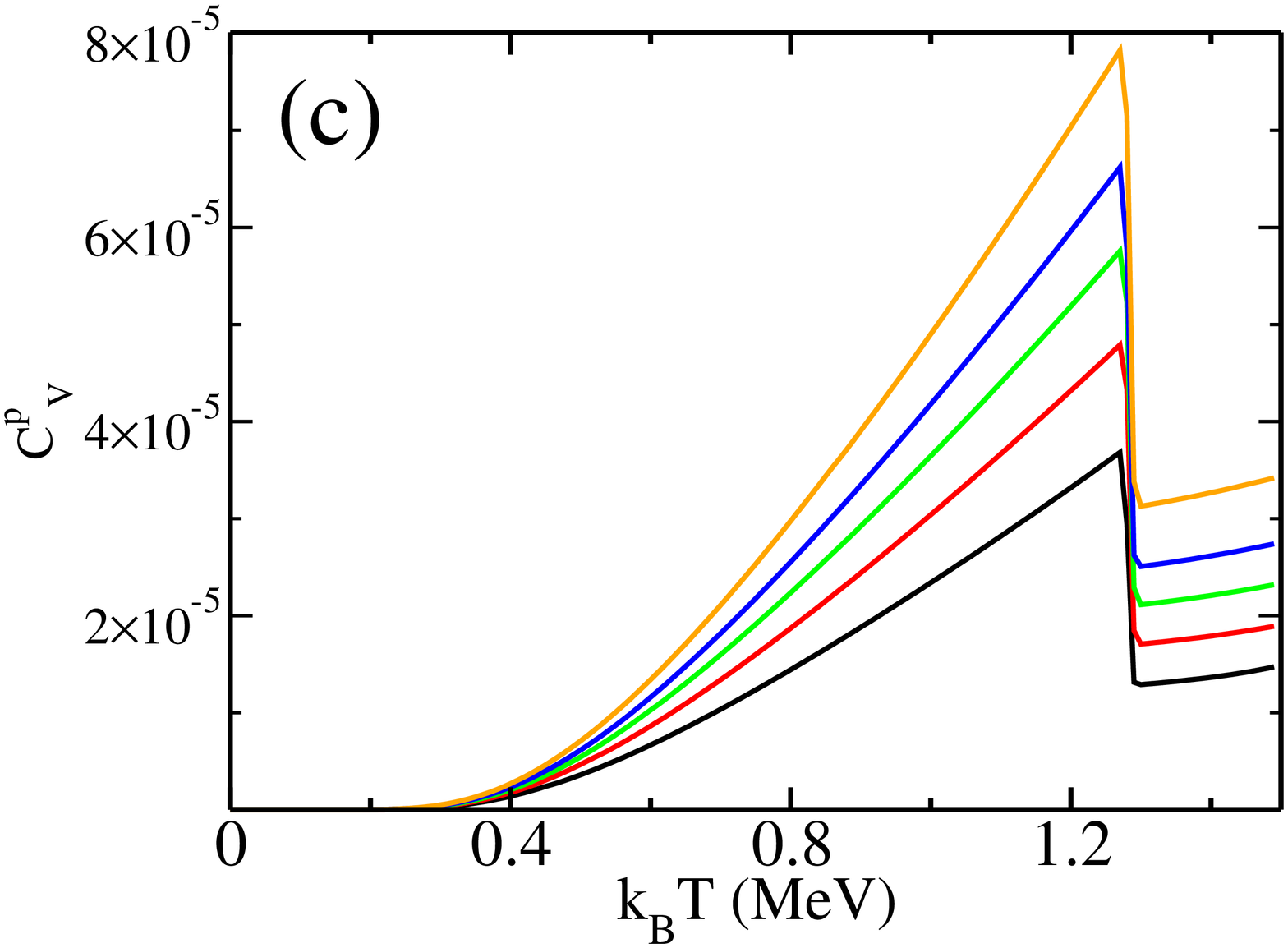}
\includegraphics[clip=,angle=-90,width=0.45\textwidth]{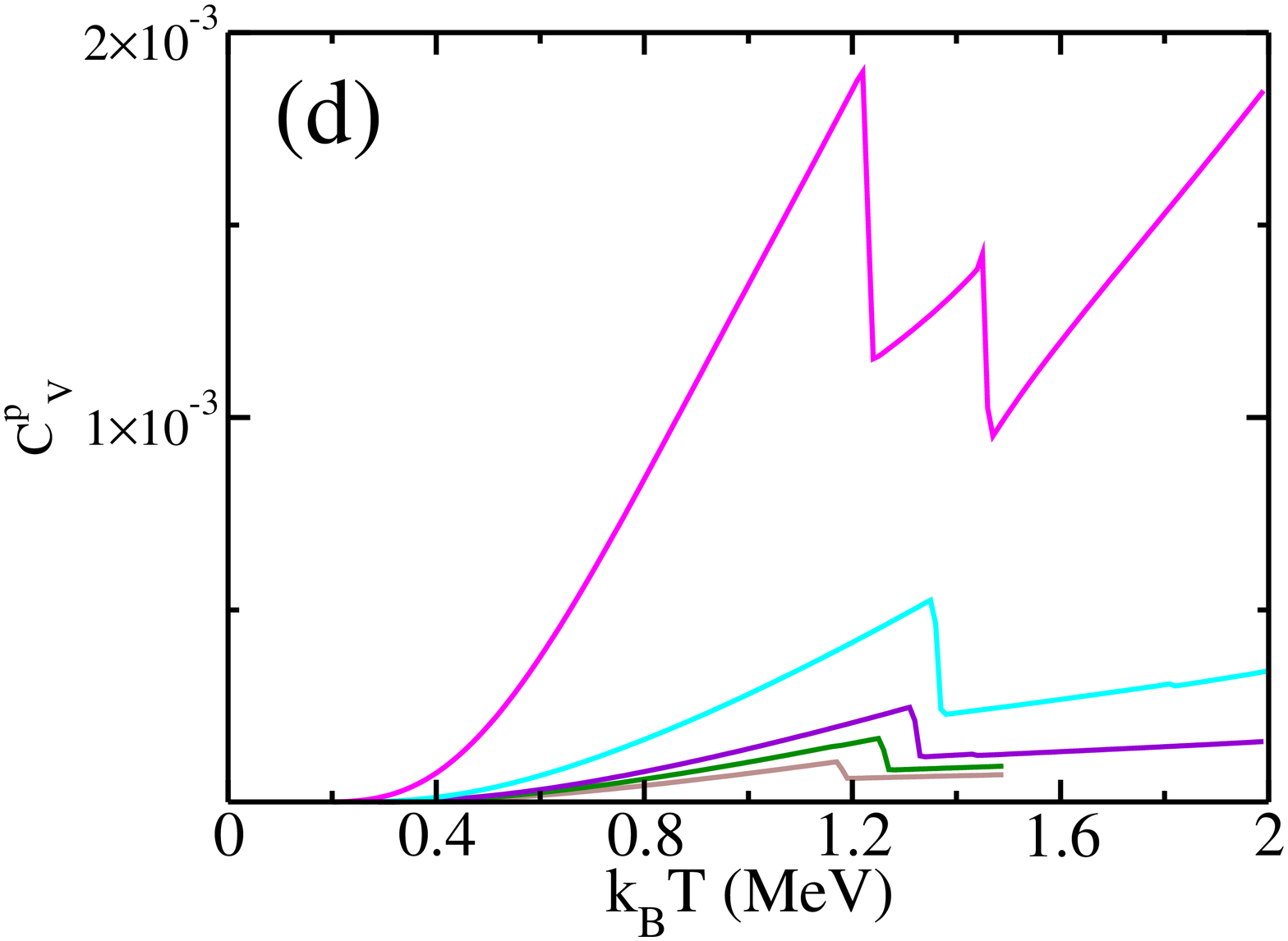}
\end{center}
\caption{
(Colors online)We show the neutron specific heat $C_{v}^{n}$ as a function of the temperature for the different cells of Tab.\ref{tabWS} obtained with Gogny D1 pairing interaction, on the bottom the specific heat $C_{v}^{p}$ for protons. See text for details.
}
\label{CVTgognyd1}
\end{figure*}

In Fig.\ref{CVTgognyd1} we show  the results of the specific heat for both protons and neutrons  obtained using the Gogny D1 interaction.
These results should be compared with the ones in Fig.\ref{CVT} for the case of $V_{\text{low k}}$ pairing interaction.
As a general remark, we observe that 
compared to the $V_{\text{low-k}}$ or DDDI results, we have the same qualitative results.
In more detail, for the cells from 11 to 8 the  specific heat is quite similar in the two calculations: in fact in this density region, the two forces give similar values of neutron paring gap, see for example Fig.\ref{Tcrit}.
For higher density cells, the difference among the critical densities obtained with the Gogny D1 interaction and the $V_{\text{low-k}}$ increases. In $^{500}$Zr the two interactions give the same critical temperature $k_{B}T_{c-\text{Gogny D1}}^{n}=k_{B}T_{c-V_{\text{low k}}}^{n}$,  while in $^{1800}$Sn we have $k_{B}T_{c-\text{Gogny D1}}^{n}-k_{B}T_{c-V_{\text{low k}}}^{n}=0.34$ MeV.
This difference in the critical temperatures is simply due to a difference in the pairing strength   and it can be understood using Eq.\ref{empiric} and comparing the values of the gaps at zero temperature for the two interactions in Fig. 5 of article I.

It is interesting to observe that, even for this interaction, we have a two peak structure in the neutron specific heat for the cells from  $^{180}$Zr to $^{500}$Zr. 
Concerning the proton specific heat, we have a very interesting result for the cell $^{1500}$Zr.
In this case we have  a clear two peak structure is shown in Fig.\ref{CVTgognyd1} (d). As already discussed in the text, we observe that the position of the second peak, corresponds to the position of a peak in the neutron specific heat in Fig.\ref{CVTgognyd1} (b).
Moreover, this is the only  case  where $C^{p}_{V}$ is only $\approx \frac{1}{10}C^{n}_{V}$, when usually the ratio among the two is much smaller $\approx\frac{1}{100}$.
We  leave the  investigation of such cell for a future work, in fact according to the results of article I, it could be that $^{1500}$Zr is stable at $k_{B}T=0$ Mev and it becomes unstable (see discussion about Fig.2 of article I)  when the pairing gap vanishes around the critical temperature.

\section{Pairing reentrance}\label{Pairsn176}

In this Appendix we shortly discuss the phenomenon of the pairing reentrance, recently discussed by Margueron \emph{et al.}~\cite{Margueron12}.
To have a better comparison with the existing results already present in the literature, we perform calculations of $^{176}$Sn using the SLy4 functional and the $V_{\text{low-k}}$ pairing interaction. We define the pairing gap as 

\begin{equation}\label{gapuv}
\Delta^{q}_{uv}=\frac{\sum_{n'nlj}(2j+1)\Delta^{q}_{nn'lj}U^{q}_{nlj}V_{n'lj}^{q}}{\sum_{n'nlj}(2j+1)U^{q}_{nlj}V_{n'lj}^{q}}.
\end{equation}

\begin{figure}
\begin{center}
\includegraphics[clip=,angle=-90,width=0.45\textwidth]{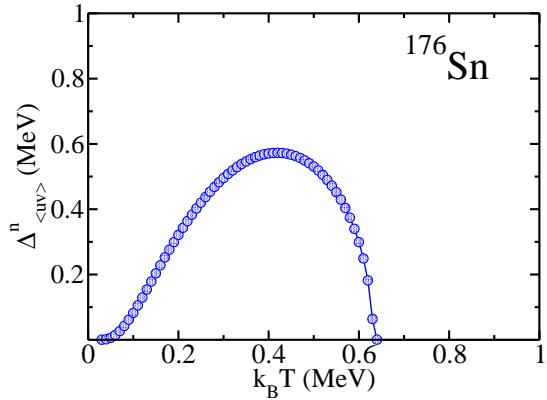}
\end{center}
\caption{
(Colors online) Averaged neutron pairing gap Eq.\ref{gapuv} for $^{176}$Sn as a function of the temperature $k_{B}T$.}
\label{sn176}
\end{figure}

\noindent We observe from Fig.\ref{sn176}, that  at $k_{B}T=0$ MeV, the system is non superfluid, then at $k_{B}T^{n}_{c}=0.05$ MeV we have a first phase transition: the nucleus becomes superfluid in neutrons. The maximum value of the gap is reached around $k_{B}T\approx 0.45$ MeV and then at $k_{B}T^{n}_{c}=0.62$ MeV, the system return non-superfluid.
This is quite different behavior compared to standard nuclei as see in ref.\cite{giai} or in our WS calculations shown in Fig.~\ref{gapDLCS}.

\noindent As discussed in the text, the Fermi-Dirac distribution, induced by the temperature, can provoke the rearrangement of the occupation of single particle levels around the Fermi energy.
In this case, some particles are moved from the close shell to some resonant states in the continuum, thus allowing the possibility to turn on the pairing field in the last shell that is no more completely occupied as it was at zero temperature.
We refer to Margueron \emph{et al.} ~\cite{Margueron12} for a more detailed analysis of such results.
As a last remark, we underlined that for the case of an isolated nucleus, it would be important to project on the good particle number.
Unfortunately, such method can not be used with most of the existing Skyrme functionals, we refer to ref.\cite{Lacroix1,Lacroix,Lacroix3} for a more detailed discussion on this problem.
We leave to a future work, a better analysis concerning the phenomenon of pairing reentrance.

\section{Finite size effects}\label{app:finitesize}

\begin{figure}
\begin{center}
\includegraphics[clip=,angle=-90,width=0.45\textwidth]{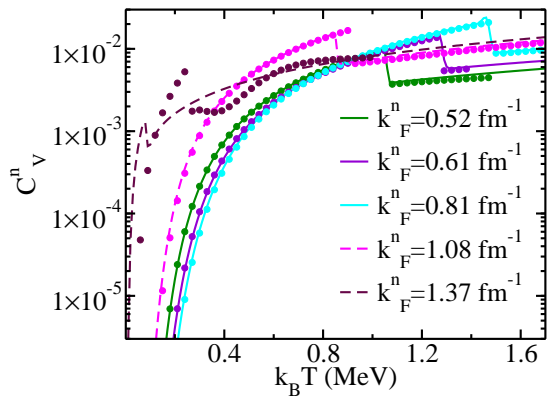}
\end{center}
\caption{
(Colors online) Neutron specific heat, $C^{n}_{V}$ (in semi-logharitmic scale), as a function of the temperature for different WS cells without protons. The calculations have been done (i) solving the HFB equations of Eq.\ref{gapeqINM} (lines) and the (ii) HFB equations in a box (dots), Eq.\ref{HFBeq}. See text for details.}
\label{finitesize}
\end{figure}

\noindent We now shortly discuss how the discretization induced by the box affects our results.
Since the main observable we have calculated in this article is the specific heat, we decided thus to compare the specific heat for the PNM case into different boxes of different size and compare the results with the one obtained by solving the HFB equations in the continuum (see discussion in Sec.\ref{pureN}).
We thus calculate here different WS cells without protons.
The value of the specific heat $C^{n}_{V}$ should not depend on the size of the box, so if we observe differences among the two methods of doing the calculations, this should originate from the discretization imposed by the box, similarly as it has been done in ref.~\cite{margue:conf09}.
The calculations have been done using the Skyrme SLy4 and $V_{\text{low-k}}$ pairing interaction.
In Fig.\ref{finitesize} we compare the specific heat of neutrons for two types of calculations: (i) we solve the HFB equations given in  Eq.\ref{gapeqINM} for a  given value neutron Fermi momentum, $k_{F}^{n}$ as given in  Tab.\ref{tabWS}; (ii) we solve the HFB equations in a box, Eq.\ref{HFBeq}, for a fixed value of $R_{WS}$ without protons and using a number of neutron that reproduces the same value of $k_{F}^{n}$ in the given box, as described in Tab.\ref{tabWS}.
We observe that the two calculations are in good agreement for boxes bigger of $\approx25$fm. When we use smaller boxes we observe the first difference due to discretization, it is the case of the cell $R_{WS}=19.6$ fm and $k_{F}^{n}=1.08$~fm$^{-1}$, where we have a small difference among the critical temperatures calculated with the two methods of about $\approx$ 80 KeV.
Very different is the case $R_{WS}=14.4$ fm and $k_{F}^{n}=1.37$~fm$^{-1}$.  We observe that the behavior of the neutron specific heat is very different in the two cases, and we restore the agreement among the two calculations only in the limit of relatively high temperatures.
Such calculations show clearly the limits of the WS approximations for high density cells.
For such kind of calculations we also tested for the cell $R_{WS}=27.6$ fm and $k_{F}^{n}=0.81$~fm$^{-1}$  the dependence of our results on the different boundary conditions when we solve Eq.\ref{HFBeq} in a box.
In such case we observe that using the BCO or BCE does not affect the results on the specific heat and on the critical temperature.

\end{appendix}

\end{document}